\documentclass[10pt]{iopart}
\usepackage{graphicx}	
\usepackage{tikz}
\usepackage[symbol]{footmisc}

\begin{document}

\title[The ARI-L project]{The Additional Representative Images for Legacy (ARI-L) project for the ALMA Science Archive}

\author{M. Massardi$^{1, 2}$,
F. Stoehr$^{3}$,
G.~J. Bendo$^{4}$,
M. Bonato$^{1}$,
J. Brand$^{1}$,
V. Galluzzi$^{5}$,
F. Guglielmetti$^{3}$,
E. Liuzzo$^{1}$,
N. Marchili$^{1}$,
A.~M.~S. Richards$^{4}$,
K.~L.~J. Rygl$^{1}$,
F. Bedosti$^{1}$,
A. Giannetti$^{1}$,
M. Stagni$^{1}$,
C. Knapic$^{5}$,
M. Sponza$^{5}$,
G.~A. Fuller$^{4}$,
T.~W.~B. Muxlow$^{4}$
}

\address{$^{1}$INAF - Istituto di Radioastronomia - Italian ALMA Regional Centre, via Gobetti 101, 40129 Bologna, Italy\\
$^{2}$SISSA, Via Bonomea 265, 34136 Trieste, Italy\\
$^{3}$European Southern Observatory (ESO), Karl-Schwarzschild-Str. 2,
85748 Garching bei München, Germany\\
$^{4}$UK ALMA Regional Centre Node, Jodrell Bank Centre for Astrophysics, Department of Physics and Astronomy,\\ The University of Manchester, Oxford Road, Manchester M13 9PL, UK\\
$^{5}$INAF-Osservatorio Astronomico di Trieste - Italian Astronomical Archives, via Tiepolo 11,34131 Trieste, Italy\\
}
\ead{massardi@ira.inaf.it}
\vspace{10pt}

\begin{abstract}
The Additional Representative Images for Legacy (ARI-L) project is a European Development project for ALMA Upgrade approved by the Joint ALMA Observatory (JAO) and the European Southern Observatory (ESO), started in June 2019. It aims to increase the legacy value of the ALMA Science Archive (ASA) by bringing the reduction level of ALMA data from Cycles 2-4 close to that of data from more recent Cycles processed for imaging with the ALMA Pipeline. As of mid-2021 more than 150000 images have been returned to the ASA for public use. At its completion in 2022, the project will have provided enhanced products for at least 70\% of the observational data from Cycles 2-4 processable with the ALMA Pipeline. In this paper we present the project rationale, its implementation, and the new opportunities offered to ASA users by the ARI-L products. The ARI-L cubes and images complement the much limited number of archival image products generated during the data quality assurance stages (QA2), which cover only a small fraction of the available data for those Cycles. ARI-L imaging products are highly relevant for many science cases and significantly enhance the possibilities for exploiting archival data. Indeed, ARI-L products facilitate archive access and data usage for science purposes even for non-expert data miners, provide a homogeneous view of all data for better dataset comparisons and download selections, make the archive more accessible to visualization and analysis tools, and enable the generation of preview images and plots similar to those possible for subsequent Cycles.
\end{abstract}

%
\vspace{2pc}
\ioptwocol

\section{Introduction: ARI-L project goals and context}

Most of the scientific results of radio interferometry lie in the continuum images and spectral line data cubes that are created by transforming the observed visibility data (the observed brightness at a given angular scale) from the Fourier domain to the image domain. However, the resulting products are not unique. In the imaging process, resolution can, for example, be traded-off against sensitivity by applying different visibility weighting schemes. The possibility to obtain more than one valid image for any visibility dataset has long justified the fact that calibrated visibilities have been considered the final product of an observation. Typically, raw data and, in some cases like ALMA, calibration scripts were stored and made available through the archive interfaces for any user to reprocess them and produce images.

The amount of observational astronomical data is growing exponentially. For example, at ESO, the data-growth is approximately 37\% per year (Stoehr 2019). Radio astronomy, with facilities like LOFAR, JVLA, MWA and - of course - ALMA, is one of the main drivers of the worldwide growth of astronomical data. Even larger amounts of data are expected to arrive when the full SKA becomes operational.

Astronomy, and in particular radio astronomy, has arrived in the era of 'big data' where the quantity of observational data is too large for astronomers to analyse it without major assistance from data reduction software and full data reduction pipelines. Consequently, in addition to significant investment in user support through three ALMA Regional Centres (ARCs), ALMA's design from the very beginning included a CASA-based (McMullin et al. 2007) data reduction pipeline\footnote{https://almascience.org/processing/science-pipeline} that would automatically perform the data calibration as well as the imaging step.

ALMA was the first radio astronomy facility to pledge that by the time the full Science Operations stage is reached, the fundamental data product of the Joint ALMA Observatory will be calibrated, deconvolved images and data cubes. These data products are not unique because of the relatively large freedom in parameter choices during the imaging process, but even in their generic form, they provide a quick way for users to assess the data-quality, the content of the data products and the interesting spatial and spectral regions. Depending on the science case, the data products may also be used for scientific analysis. For these reasons, the image products are delivered to the ALMA users through the ALMA Science Archive (ASA).

From ALMA's Early Science period up to Cycle 3, the imaging-dedicated part of the ALMA pipeline was not available. The staff at the observatory and at the ARCs manually performed the quality assessment (QA2) of the data before they were delivered to the PIs (Petry et al. 2020). This manual procedure was carried out for over 1800 ALMA projects. The manual imaging of each full data set is very time consuming, so the analysis often focused on only a small subset of a project's calibrated data that was just large enough to assess the quality. As a result, only a very small fraction of all raw data ($< 10\%$) was converted into images and image cubes. This fraction increased drastically from Cycle 5 (i.e. from 2017) onwards when the ALMA Science Pipeline was used nearly exclusively for QA2 data reduction.

Use of the archive has increased over the years, and the ASA is now considered an important, widely-utilized resource; about 25\% of all ALMA publications are based either purely on archival data or use archival data in addition to PI data (Stoehr et al. 2017).

The availability of deconvolved images and data cubes vastly speeds up researchers' data analysis process. Rather than having to download the raw data, to identify the CASA versions to use, to run the calibration or restoration script, and to modify and run the imaging script for hours or days just to determine whether objects or spectral lines have been detected, researchers can use previously-created products to make these types of assessments in seconds.

In addition to allowing direct analysis by archival researchers, these products also enable additional services. This includes the visualisation of products using the remote visualisation tool CARTA\footnote{https://cartavis.github.io/}, the creation of preview images that can be displayed directly on the ALMA Science Archive query interface\footnote{https://almascience.org/asax}, the automated post-analysis of the ALMA data products (for example, with the ALMA Data mining ToolkIT ADMIT\footnote{http://admit.astro.umd.edu/}, Teuben et al. 2015, or the Keywords of Astronomical FITS-Images Explorer KAFE, Burkutean et al. 2018), and the ability to directly access the products through VO services like TAP, SIA2 and DataLink, that allow for spectral multi-band comparisons, source identifications, and catalogue reconstruction.

The Additional Representative Images for Legacy (ARI-L) project is an ALMA Development project approved by JAO and ESO in December 2018 following an ALMA development study approved by ESO and reported by Massardi et al. (2019). The aim of ARI-L, as presented in this paper, is to use the ALMA Imaging Pipeline (introduced in 2017) on the data from early observing Cycles (2-4) so as to create, where possible, data products of the same completeness and quality as ALMA is creating for new observations nowadays and to ingest them in the ASA.

Over its three years of activity, ARI-L will produce and ingest into the ASA a uniform set of full data cubes and continuum images that cover at least 70\% of the data from Cycles 2-4 that can be processed with the ALMA Pipeline. These cubes complement the existing QA2-generated image products.

The ARI-L project activity started in June 2019. The first data products were delivered to JAO in November 2019, and the project has been ongoing since then and has produced, assessed the quality of, and delivered more than 150000 images that can already be downloaded from the ASA. ARI-L products are certified for quality and ingested into the archive as so-called ``external products''. The ARI-L project documentation and website as well as periodic updates are publicly available through the ALMA science portal \footnote{ https://almascience.org/alma-data/aril}.

In this paper, we present the project principles, the workflow, and the data products (see Sect. \ref{sec:workflow}).  We then discuss some new opportunities offered to the ALMA Science Archive users through ARI-L products (in Sect. \ref{sec:status}), and we finally (in Sect. \ref{sec:summary}) summarize the current and future project plans.

\begin{table}
\caption{\label{tab:arch_stats} Number and properties of Member Observing Unit Sets (MOUS) for ALMA Cycles 2-4 that can be processed in the ARI-L project, including QA2-passed and -semipassed projects but excluding full-Stokes, VLBI, solar, and TP observations as well as projects that have already been processed with the Imaging Pipeline and are not in QA3 (see the footnote for definition) as of June 2021.}
\footnotesize
\begin{tabular}{l|ccc|c}
\br
Cycle& 2&3 &4 & Total\\
\mr
ARI-L MOUS & 1085 & 1719 & 672& 3476 \\
\% of observed MOUS & 82.3 & 83.5 & 32.0 & 63.5\\
Total raw data size [TB] & 22 & 66 & 50 & 138 \\
Median size per MOUS [GB]&11.2&16.8&20.3 & 14.7 \\
\br
\end{tabular}
\end{table}
\normalsize

\section{The ARI-L Project workflow}\label{sec:workflow}

\subsection{ARI-L terms of reference}\label{sec:principles}

ALMA observations and products are organized hierarchically. Each project is identified by a code with the form of `YYYY.C.nnnnn.T' in which `YYYY' corresponds to the year in which the project was submitted, `C' corresponds to the call, `nnnnn' is the sequential number assigned to the proposal when it was submitted, and `T' is the proposal type. ARI-L only works on data from Cycles 2, 3, and 4, i.e. project IDs that start with '2013', '2015' and '2016'.

Cycles 0 and 1 only make up 2\% and 8\% of the raw data size of the first 5 ALMA Cycles, respectively. The Cycle 0 and 1 data format and packaging have differences with respect to the following Cycles and therefore require dedicated procedures. However, most of the objects have been re-observed in later Cycles using more antennas. Our study (Massardi et al. 2019) indicates that it is preferable to focus on more recent Cycles. The Imaging Pipeline has been used for ALMA products since early 2017 (during Cycle 4), so the products from subsequent Cycles are already as complete as the pipeline processing allows.

PIs specify their observations as a series of Science Goals. At operation stage, this corresponds to one or more `Group Observing Unit Sets', each of which may be split in various `Member Observing Unit Sets' (MOUSs) that include the instrumental setting needed to reach the PI's goals. Each MOUS translates into `Scheduling Block' commands during observations that, for calibration purposes, are limited in execution time. Hence, multiple repetitions (or `Execution Blocks', which can be abbreviated as EBs) might be necessary to reach the requested sensitivity. Each EB is calibrated individually. Because calibrated data could be combined to generate images at the MOUS level, the selection and analysis level for ARI-L is the MOUS. Table \ref{tab:arch_stats} lists the number of MOUSs that are being processed in the ARI-L project.

Only datasets that are accessible for public download and that have passed ALMA quality control (either 'PASS' or 'SEMIPASS' in QA2\footnote{An MOUS is considered 'PASS' in  level 2 of the Quality Assurance procedure if after the full calibration and generation of imaging products it reaches the conditions of sensitivity and resolution requested by the PI in the proposal that generated that dataset. 'SEMIPASS' indicates that the dataset falls short of meeting the PI-requested sensitivity and resolution, but are otherwise of good quality (see ALMA Technical Handbook).}) are selected for ARI-L processing. This excludes data which are in QA3\footnote{Level 3 of the Quality Assurance could be initiated by PIs in case of issues encountered after product delivery. QA3 typically reactivates data access restrictions for MOUSs, allowing only PIs and users with delegated access to download the data.} and data still covered by a proprietary period.

Furthermore, we select only datasets with observing modes that can be handled by the imaging tasks of the ALMA Pipeline (hereafter defined as `Imaging Pipeline'). Many of the observing modes that were considered non-standard when observed (and thus required special care during calibration) can now be processed by the Imaging Pipeline as common standard observing modes.  However, solar, full-Stokes, and VLBI observations are not reprocessed because they cannot be handled by the Imaging Pipeline. Similarly, Total Power datasets are also not reprocessed because the full data reduction (calibration and imaging) for those data has already been completed.

The ARI-L products are intended for archive-mining purposes and as an overview of the archived data content. They are not necessarily science-ready. However, the ARI-L project also uses the Imaging Pipeline outside the range of goals for which it has been commissioned, as it is the best tool currently available to create homogeneous images of data from past Cycles.

For processing purposes, we sort the selected MOUS as an increasing function of the expected run time of the datasets, processing the faster ones first. Our tests (Massardi et al. 2019) demonstrated that a very rough run-time score value for an MOUS can be constructed using a function of its data volume and the maximum distance between the antennas during the observations (the maximum baseline). The data volume is related to the number of EBs (defined by the duration of the observing run and by the presence of multiple pointings, which could be either in a mosaic or covering multiple targets) and to the number of spectral channels; more time is needed to create images with larger data volumes. The maximum distance between the antennas during the observations is an indication of the resolution that can be achieved in the image, and more time is needed to create higher-resolution images.

Processing parameters are typically applied in the same way as they are in later Cycles.  A README file is provided for each imaged MOUS  summarising the data processing and any deviations from standard procedures.

  \begin{figure*}
  \centering
  \includegraphics[trim={2cm 2cm 2cm 0cm,clip},width=0.8\textwidth ]{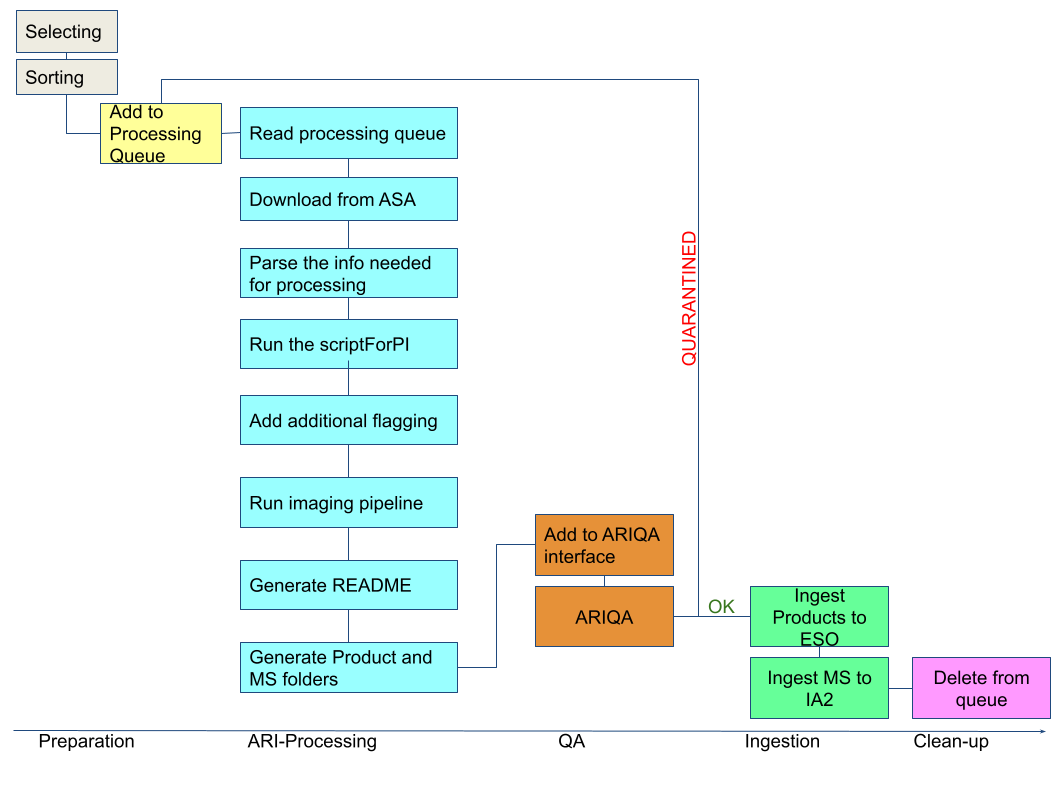}
  \caption{The decision tree and the different phases of the ARI-L project as applied to each MOUS in Cycles 2-4.}
      \label{fig:process}
  \end{figure*}

\subsection{ARI-L processing}\label{sec:processing}

The ARI-L products are generated by a Python-based workflow engine (the ARI-L code). According to the above-stated terms of reference, for each processable MOUS, the ARI-L code

\begin{itemize}
 \item downloads the raw data, auxiliary data (including the calibration scripts and tables), and existing products from the ASA;
 \item identifies the original CASA version used to generate the calibration scripts that are stored in the ASA;
 \item runs the scriptForPI to restore the calibrated Measurement Sets (MS);
 \item adds additional flagging that was identified by the original QA2 data-reducers;
 \item uses the latest CASA version available at the start of processing (so far CASA 5.6.1-8) to generate the data-products with the Imaging Pipeline, extending the processing to also include data-cubes for calibrator observations;
 \item packs the calibrated MS for permanent storage;
 \item renames the products to follow the ALMA naming standard;
 \item compresses the auxiliary files;
 \item generates a README file with a summary of the content of the MOUS;
 \item runs an automatic comparison with the existing manually created QA2 products (see Section 2.5);
 \item catches processing errors and identifies classes of errors for later recovery;
 \item stores the processing information in a relational database system.
\end{itemize}
After the ARI-L code processing, the data undergo the ARI-L quality assurance step before they are delivered to ESO and then JAO for ingestion into the ASA. The ingested files are cross-checked with the generated products, and, if they match, the latter are deleted from the ARI-L processing area. A visual representation is shown in Fig.\,\ref{fig:process}

\subsection{Restoring calibration}

Packages of archived datasets include all the scripts used during the ALMA Quality Assurance processes for data reduction and for checking whether the sensitivity and resolution required by the PI have been reached. In the early Cycles, all the MOUS were manually calibrated by experts from the ALMA Regional Centres using scripts for calibration and imaging generated from a standard template.  Despite strictly following common checklists that guarantee the quality of the final products, each analyst was free to modify the scripts as they felt was necessary. For this reason, early Cycle scripts might differ from each other.

The information needed for the ARI-L processing (e.g. the CASA version needed for calibration) is retrieved automatically from the downloaded data. Then, for the execution blocks (EB) of each MOUS, calibration is performed with an appropriate CASA version using the prescriptions of the calibration scripts included in the downloaded folder. No change or verification is applied to the calibration stage except on rare occasions where problems had been identified in early procedures.

As mentioned above, imaging scripts typically are written to produce example images or data cubes that demonstrate that the requirements of the PI were achieved in the observations. Hence, the QA2-produced imaging scripts included in the script folder are not used except for extracting meaningful additional flagging commands. This may happen in manual imaging when the QA2 analyst realized that, at the end of the calibration, a portion of the data (i.e. some channels, a time range, one or more antennas, some baselines, etc.) are still significant outliers with respect to all other data and therefore decided to flag the outlying data after the calibration but before imaging. After these flagging commands are extracted, they are applied in the ARI-L procedure after the calibration and before the imaging. This process generates the calibrated measurement sets that we give as input to the Imaging Pipeline.

\subsection{Imaging with the ALMA pipeline}

The calibrated measurement sets are processed with the most recent version of the Imaging Pipeline that has been tested for ARI-L purposes. In the first year of ARI-L activities, we have used the version included in the CASA release 5.6.1-8\footnote{We refer to the ALMA Science Pipeline User's Guide (https://almascience.org/processing/documents-and-tools/alma-science-pipeline-users-guide-casa-5-6.1) for details on the parameters used.}. The Pipeline task sequence is the same as that implemented in later Cycles.

Imaging is performed for each source included in each MOUS.  We image not only the science targets but also the bandpass, phase, and check source calibrators. For each target and calibrator in each MOUS, the ARI-L procedure, after uv continuum subtraction, produces aggregate continuum images and multi-frequency synthesis (mfs) continuum images and spectral cubes for each spectral window. We take advantage of the improved continuum-finding abilities of the recent pipeline to detect weak line emission, particularly in ACA 7-m data. Consequently, the estimated noise per channel is ameliorated with respect to previous CASA versions, allowing for a more robust S/N thresholds derivation for the associated channels. The improved automasking is beneficial for both emission and absorption line detection. Nonetheless, this amelioration may still be inadequate in the case of weak calibrators; see the CASA guide\footnote{https://casaguides.nrao.edu/index.php/Automasking\_Guide} for more information.

All ARI-L images are primary beam corrected. The corrected images cover a region of diameter equal to the FWHM of the primary beam centered at the pointing position. The Imaging Pipeline by default attempts automasking of the images. Primary beam and mask maps are delivered together with the product image.

For cubes, the channel resolution is defined by the native resolution of the observations but also accounts for the nominal channel size having been reduced by a factor $\sim 2$ by Hanning smoothing. We allowed the Imaging Pipeline to set the optimal parameters and masks for each image.

The Briggs robust parameter is set to 0.5 for all data unless the Imaging Pipeline indicates a different approach. In all cases, the robust parameter is stored in the image header in the ROBUST keyword so that the applied value can always be verified.

In case the size of the product file is larger than 180~GB, the Imaging Pipeline has implemented a mitigation process by which only one-third of the primary beam is imaged or the image resolution is smoothed to 3 pixels per synthesized beam. In later Cycles, channel averaging might also be applied to high spectral resolution data that produce cubes that are too large or to data from mosaic observations where the size of the field made it impossible to produce final images. In extreme cases, only a few fields and/or a few spectral windows are imaged. We set a threshold size for mitigation 3 times larger than the one used in later ALMA Cycles to allow for better completeness of our products.
In ARI-L Cycles 3 and 4 processed MOUS mitigation cases constituted less than 1\%; these data were checked on a case by case basis. In case images of target sources or spectral windows are missing because of mitigation a note is added to the ARI-L README.

The Imaging Pipeline produces a Weblog report that is delivered in the ARI-L product folder and reports all the commands, parameters, and tasks used. However, this Weblog is available only for the sources observed as science targets, not for the calibrators. Every step listed in the Weblog also gets a score that indicates its success with respect to common heuristics. In case the pipeline cannot apply automasking, applies mitigation, or defines a different set of weighting parameters, a warning flag in the Weblog is raised. In case the resultant images are considered of good quality they are accepted.

This approach has been successfully tested on all the ALMA Cycles within the scope of ARI-L and used thus far to generate mostly Cycle 3 and 4 products. Some additional caveats in the usage of the Imaging Pipeline on Cycle 2 data are described in the product README files whenever the procedure differs from the one described above.

\subsection{The ARI-L Project quality assurance (ARIQA)}

The ARI-L products have a quality that is virtually always comparable to the existing manually-created products from QA2. However, ARI-L offers complete cubes for all the sources in the datasets, while the manually imaged products often show only a small fraction of the available data (eg. only one or a few sources or a limited number of channels).

All the ARI-L products are quality-checked by our team of expert data-analysts before being ingested into the archive. The Additional Representative Images Quality Assurance (ARIQA) procedure has four layers.
\begin{itemize}
\item A check is performed to ensure that the ARI-L code was executed correctly. In case of failures, ARI-L products will not be ingested into the ASA.
\item Another check is performed to verify that all the images that were expected are produced. The causes of missing images are carefully identified. Decisions about missing images are made on a case by case basis and, if necessary, reported in the README.
\item The Weblog of the Imaging Pipeline products is reviewed to verify that the pipeline has properly generated good quality images according to its heuristics (i.e. all the ALMA Imaging Pipeline tasks have been executed with a score larger than 0.90). Any discrepancy is analysed and, if the discrepancy is not justified, the relative ARI-L products are not ingested into the ASA.
\item The portion of the data corresponding to what is available as QA2 products are extracted and smoothed to the same resolution, and then the rms noise and dynamic ranges are compared. Differences should be evaluated on the basis of the processes used during QA2. In case the differences are not justified and the resulting rms or dynamic range worsen by more than 30\% or the imaged structure is clearly wrong, the ARI-L products are not ingested into the ASA.
\end{itemize}

The ARIQA is not intended to be a repetition of QA2 and will never change the QA2 outcome for an MOUS. It only evaluates the quality of the produced images and relies on the calibration for each MOUS as it is done in the archived scripts. For this reason, once the ARI-L code has successfully produced an image product, any quality comparison will be done with the existing QA2 images and not in reference to the project's requirements in sensitivity and angular resolution. Nevertheless, ARI-L images have been through a distinct and less manually-intensive QA process compared to the corresponding products generated in the ALMA QA2 process. Image characteristics resulting from these distinct processes are typically comparable but can vary, particularly when manual adjustments were required for QA2 (e.g. self-calibration or optimization of the continuum channel selection). Furthermore, we stress that the ARIQA process ensures reliable imaging products for manually calibrated data sets as well.

In cases where only some of the images of an MOUS can pass the ARIQA, we may consider ingesting only those that passed ARIQA into the ASA even after accurate evaluations and attempts to recover the other images during the quality assurance process. If the products still have any issues after the ARIQA but their quality is considered sufficient for delivery, a note is added in the README file enclosed in the ARI-L product folder and they are delivered.

\subsection{The ARI-L products delivery and retrieval}

For all the successfully processed MOUSs that pass quality control, the final ARI-L calibrated measurement sets are stored in a dedicated storage system outside the ASA that is hosted and maintained by the INAF-IA2\footnote[1]{www.ia2.inaf.it} facility. The data are available to the user community via the ARI-L webpage. The measurement sets will be stored for the whole duration of the ARI-L project and for a minimum of 3 years after the conclusion of the project. In this system, the calibrated dataset folder, which is named after the original observing project and MOUS, includes measurement sets for each calibrated EB.

The imaging products from MOUSs that passed the ARIQA process are delivered to ESO to be ingested in the ASA. A README file is enclosed in the ARI-L product folder. Its purpose is to trace the history and hierarchy of the dataset used to generate all the product images, including:
\begin{itemize}
  \item a synthetic description of the purposes of the ARI-L products;
  \item the description of the MOUS as taken from the archived QA folder;
  \item notes on the ARI-L process (including at least the CASA version used for processing);
  \item a table describing the content of the EBs included in the MOUS (one per line): sequential number of the EB; EB name; time range in MJD of the observations; array (7-m, 12-m); number of antennas; frequency range for each spectral window in GHz; spectral resolution for each spectral window in MHz; range of angular scales covered in arcsec; first, second, and third quartile of the uv range in metres;
  \item a table summarising the properties for each source observation (science target or calibrator; one per column) included in the MOUS: right ascension and declination in degrees; sequential number of EB as in the previous table in which the source is included; number of pointings (which differs from unity in case of mosaics); largest angular scale and resolution of the observation (in arcsec); intent of the observation of the source.
\end{itemize}

Each ARI-L product file in the ASA archive includes the `ari-l' tag in the name along with indications of the original project and MOUS numbers, the spectral windows, and the source names, all following the more recent ALMA file naming scheme. ARI-L image, README and Weblog files can be retrieved from the ALMA Science Archive in the form of single files as "Externally delivered products" on the Request Handler download page; these files are directly listed below the ALMA data.

\subsection{The ARI-L failure analysis and recovery}

MOUSs that failed ARI-L processing or ARIQA are temporarily quarantined. They are then classified and analysed during dedicated recovery sessions. We consider reprocessing each MOUS using the same or different Imaging Pipeline parameters and attempt this at least once for each quarantined MOUS. If successful, a note is added to the README with the final processing parameters if they are different from the standard ones. This approach is important in guaranteeing the homogeneity of our delivered products and our control on the application of the Imaging Pipeline to old datasets, and it improves our knowledge of the data processing as well.

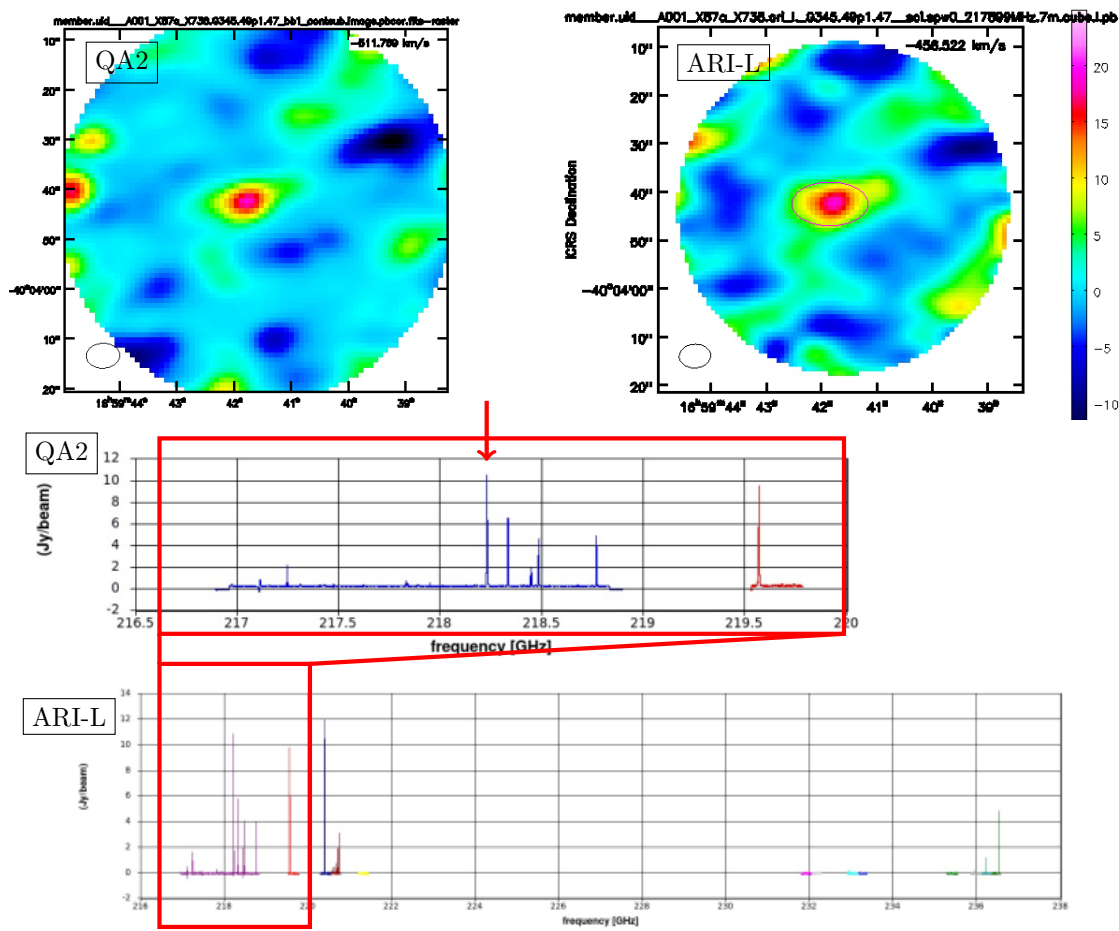
\begin{figure*}
\centering
\begin{tikzpicture}
\node (img) at (-10cm,9cm) {\includegraphics[width=6cm]{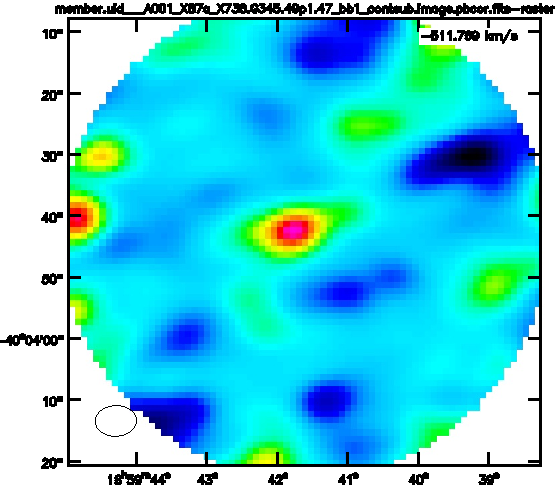}};
\node [draw] at (-11.5cm,11cm){QA2};

\node (img) at (-2cm,9cm) {\includegraphics[width=7.5cm]{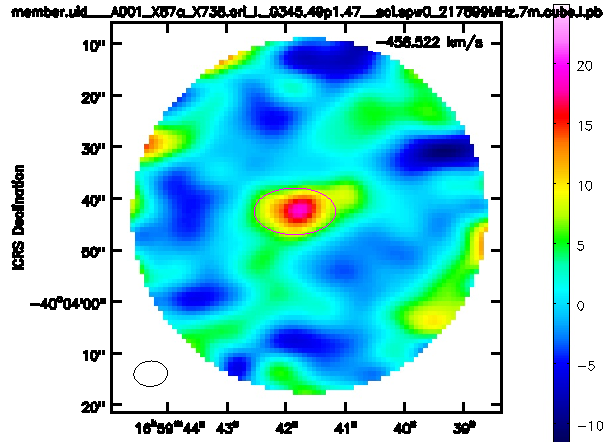}};
\node [draw] at (-3.5cm,11cm){ARI-L};

\node (img) at (-7.2cm,4.5cm) {\includegraphics[height=3cm]{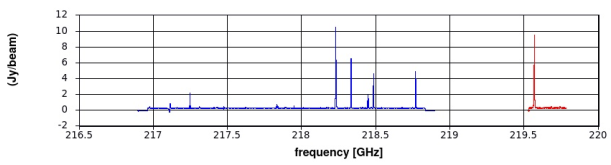}};
\node [draw] at (-12.3cm,5.8cm){QA2};
\draw [->, red, ultra thick] (-6.65cm,6.5cm) -- (-6.65cm,5.7cm);
\draw [red, ultra thick] (-11cm,6.0cm) rectangle (-1.9cm,3.4cm);

\node (img) at (-5.5,1.2cm) {\includegraphics[height=3.5cm]{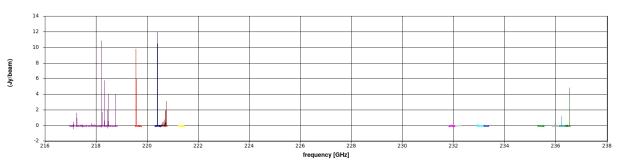}};
\node [draw] at (-12.2cm,2.3cm){ARI-L};
\draw [red, ultra thick] (-11cm,3.0cm) rectangle (-9cm,-0.5cm);
\draw [red, ultra thick] (-11cm,3.4cm) -- (-11cm,3cm);
\draw [red, ultra thick] (-1.9cm,3.4cm) -- (-9cm,3cm);

\end{tikzpicture}
\caption{Comparisons of the QA2 and ARI-L products for the source G345.49+1.47 in the MOUS uid://A001/X87a/X738 from the project 2016.1.00288.S. The top row shows images of the emission at specific channels.  The second and third rows show the complete spectra extracted from the data available in the ASA, with red arrows indicating the part of the spectrum that the images in the top row are extracted from. In the QA2 products, images without continuum subtraction are available for the first two spectral windows, and a continuum-subtracted cube is available for only the first one. In the ARI-L products, continuum-subtracted cubes are available for all 13 spectral windows, as shown in the complete ARI-L spectrum and in the next figure}.
  \label{fig:G345_1}
\end{figure*}

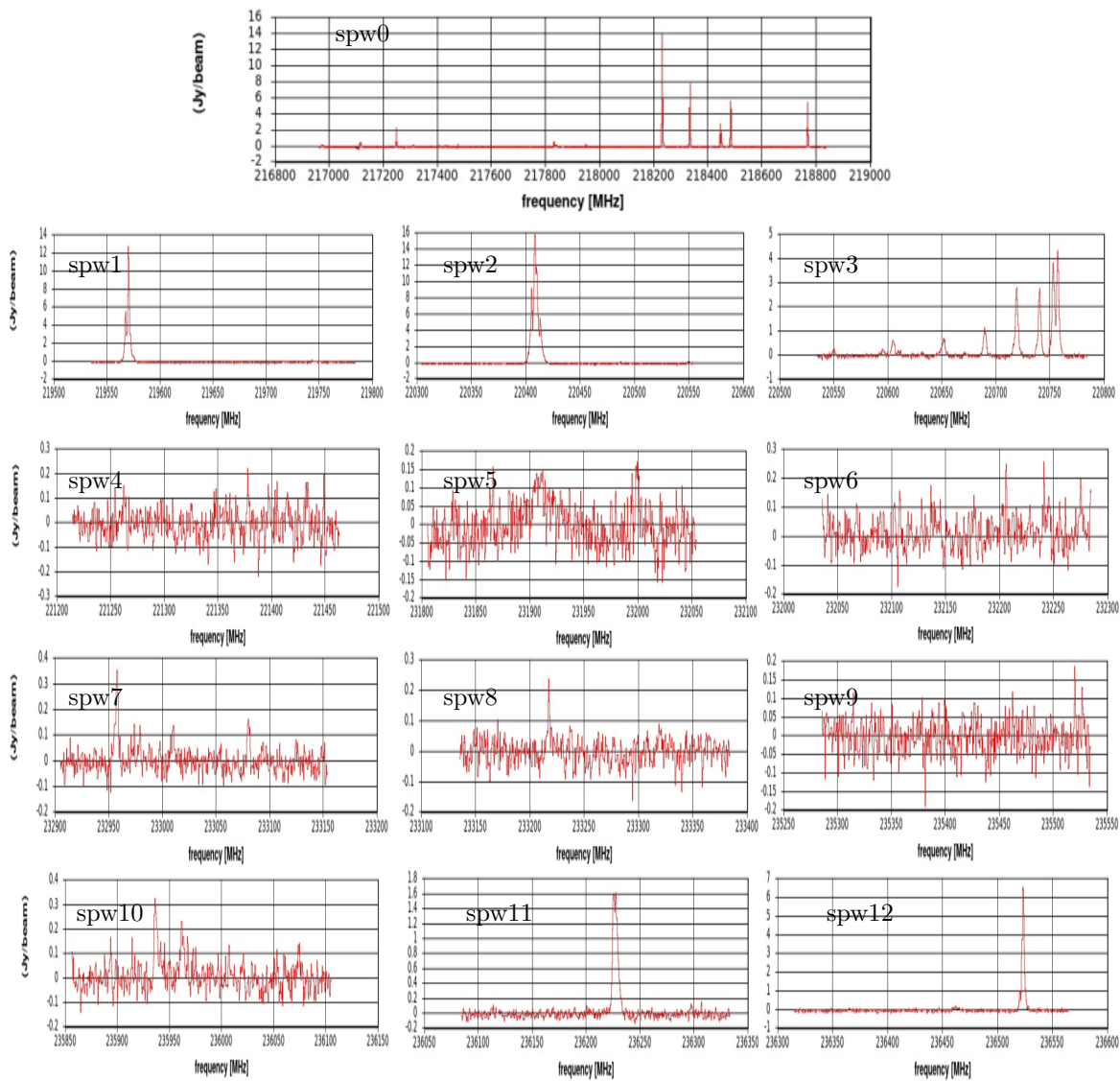
\begin{figure*}
\centering
\begin{tikzpicture}
\node (img) at (-8cm,9cm) {\includegraphics[height=3cm, width=10cm]{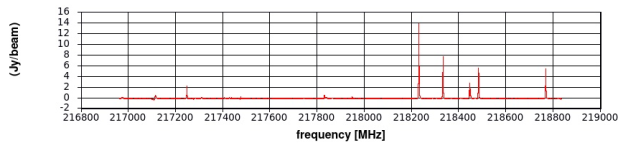}};
\node at (-10.5cm,10cm){spw0};

\node (img) at (-12.7cm,6.1cm) {\includegraphics[height=3.2cm, width=5.5cm]{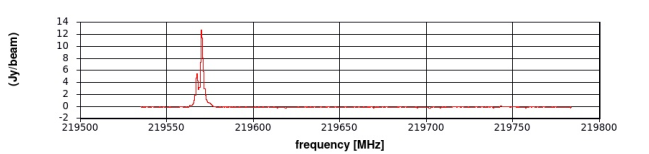}};
\node at (-14.2cm,6.8cm){spw1};
\node (img) at (-7.5cm,6cm) {\includegraphics[height=3cm, width=5cm]{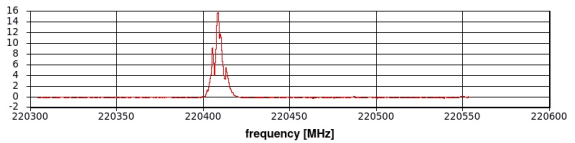}};
\node at (-9cm,6.8cm){spw2};
\node (img) at (-2.5cm,6cm) {\includegraphics[height=3cm, width=5cm]{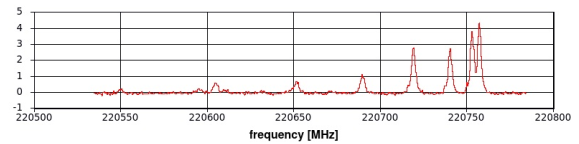}};
\node at (-4cm,6.8cm){spw3};

\node (img) at (-12.7cm,-2.9cm) {\includegraphics[height=3.1cm, width=5.2cm]{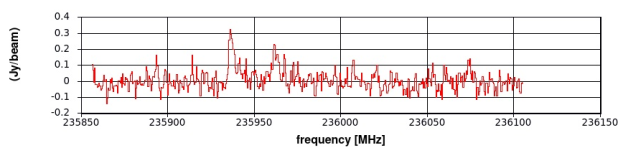}};
\node at (-14.cm,-2.2cm){spw10};
\node (img) at (-12.7cm,0.05cm) {\includegraphics[height=3.2cm, width=5.5cm]{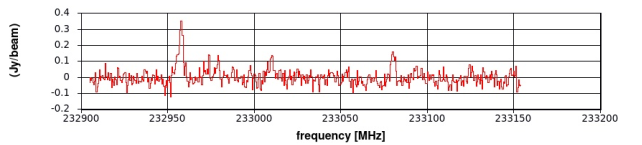}};
\node at (-14.2cm,0.8cm){spw7};
\node (img) at (-12.7cm,3cm) {\includegraphics[height=3.0cm, width=5.5cm]{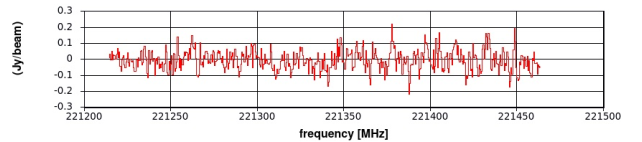}};
\node at (-14.2cm,3.8cm){spw4};

\node (img) at (-7.5cm,2.95cm) {\includegraphics[height=3cm, width=5cm]{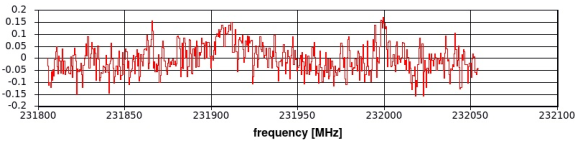}};
\node at (-9cm,3.8cm){spw5};
\node (img) at (-2.5cm,3cm) {\includegraphics[height=3.1cm, width=5cm]{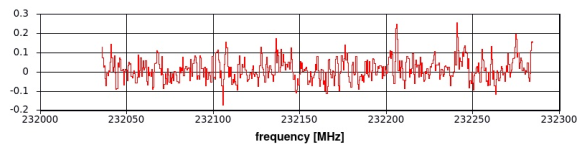}};
\node at (-4cm,3.8cm){spw6};

\node (img) at (-7.5cm,0.cm) {\includegraphics[height=3cm, width=5cm]{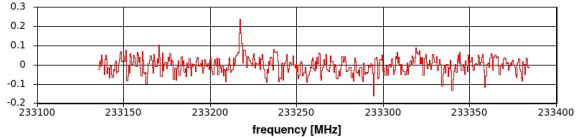}};
\node at (-9cm,0.8cm){spw8};
\node (img) at (-2.5cm,0cm) {\includegraphics[height=3cm, width=5cm]{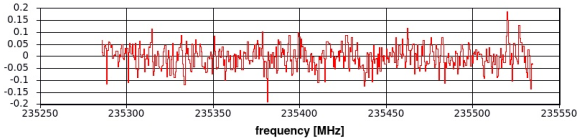}};
\node at (-4cm,0.8cm){spw9};

\node (img) at (-7.45cm,-3.0cm) {\includegraphics[height=3cm, width=4.9cm]{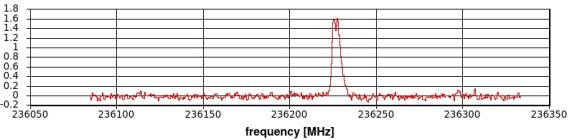}};
\node at (-8.6cm,-2.2cm){spw11};
\node (img) at (-2.5cm,-3.0cm) {\includegraphics[height=3cm, width=5cm]{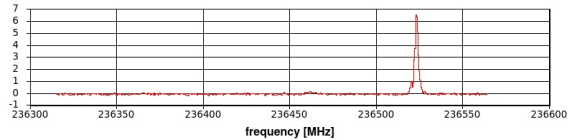}};
\node at (-3.6cm,-2.2cm){spw12};

\end{tikzpicture}
\caption{Spectral of ARI-L products for the source G345.49+1.47 in the MOUS uid://A001/X87a/X738 from the project 2016.1.00288.S, presented in the previous figure: continuum-subtracted cubes are available for all 13 spectral windows. Almost all of the ARI-L spectra  show clear spectral features.}
  \label{fig:G345_2}
\end{figure*}


\begin{figure*}
\centering
\begin{tikzpicture}

\node (img) at (-7cm,9cm) {\includegraphics[trim=4.5cm 1cm 3.2cm 0.8cm,clip, width=5.5cm]{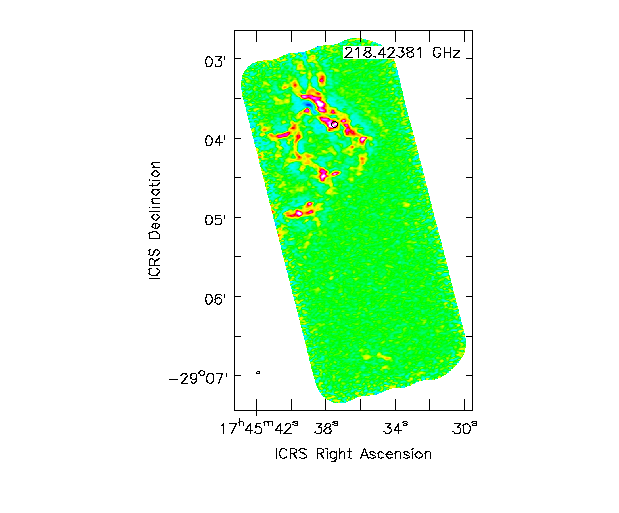}};
\node [draw, fill=white] at (-8cm,12.2cm){ARI-L};

\node (img) at (-12.5cm,9cm) {\includegraphics[trim=4.5cm 1cm 3.2cm 0.8cm,clip, width=5.5cm]{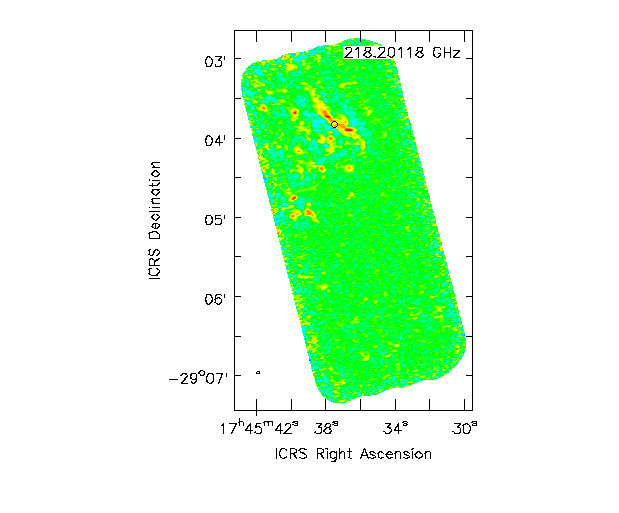}};
\node [draw, fill=white] at (-13.5cm,12.2cm){ARI-L};

\node (img) at (-18.1cm,9cm) {\includegraphics[trim=2cm 1cm 2cm 0.9cm,clip, width=6cm]{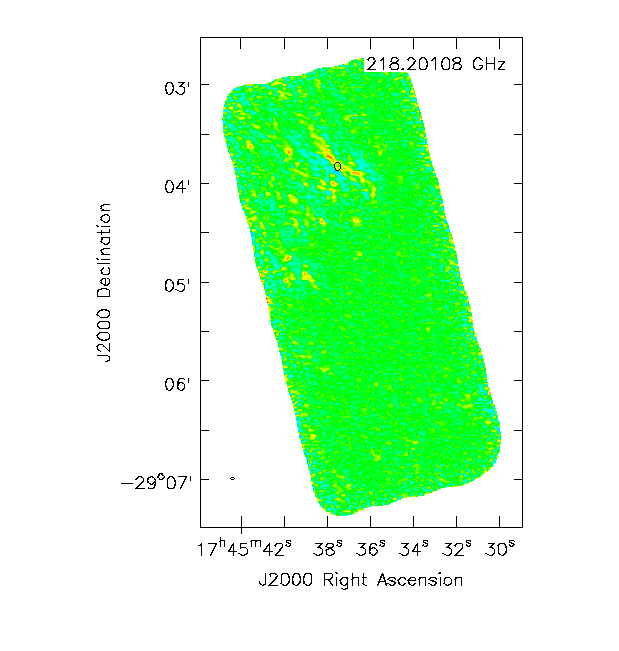}};
\node [draw, fill=white] at (-19cm,12.2cm){QA2};

\draw [green, ultra thick, dashed] (-21cm,13.0cm) rectangle (-10cm,5cm);

\draw [red, ultra thick, dotted] (-9.8cm,13.0cm) rectangle (-4.2cm,5cm);

\node (img) at (-12.5cm,3cm) {\includegraphics[trim=0cm 0cm 0cm 1cm,clip,width=\textwidth]{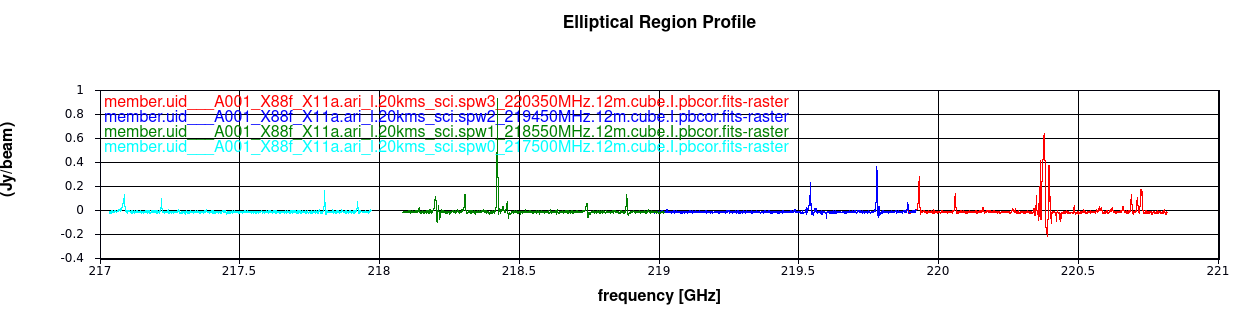}};
\node (img) at (-12.5cm,-0.6cm) {\includegraphics[trim=0cm 0cm 0cm 1cm,clip,width=\textwidth]{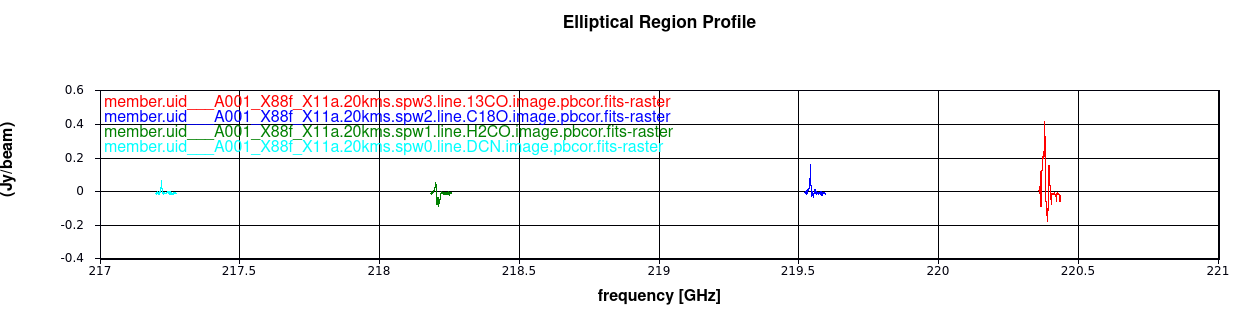}};

\node [draw, fill=white] at (-19cm,4.5cm){ARI-L};
\node [draw, fill=white] at (-19cm,1cm){QA2};

\draw [->, green, ultra thick, dashed] (-15.1cm,5cm) -- (-15.1cm,3cm);
\draw [->, red, ultra thick, dotted] (-9.7cm,5cm) -- (-14cm,3cm);

\end{tikzpicture}
\caption{Images and spectra ingested into the ASA for the project 2016.1.00875.S MOUS uid://A001/X88f/X11a (a spectral scan of a region of the Central Molecular Zone of the Milky Way). The first two images show the QA2 and ARI-L products for the same channel, which is indicated by the green arrow in the two spectra shown below. The image on the right shows the structure associated with the bright spectral feature indicated by the red arrow in the ARI-L cube spectrum; this is absent in the QA2 cubes.}
  \label{fig:comparison}
\end{figure*}

\section{Examples of ARI-L products}\label{sec:status}

The project officially started in June 2019 and is planned to finish after three years. The project deployment is on schedule, most of the expected products are already available for download from the ASA, and the relative calibrated measurement sets can be requested on our website accessible through the ALMA Science Portal\footnote{https://almascience.org/alma-data/aril}.

Given the large number of science cases involved, it is clear that a comprehensive overview of the scientific opportunities offered by the ARI-L images is not possible and also beyond the scope of the current paper. Massardi et al. (2019) list a few possible science cases. In the following, we present just a few examples of the improvements and new applications that ARI-L products allow.

 The most relevant added-value that ARI-L brings to the ASA is the completion of the images for the older Cycles. Figures \ref{fig:G345_1}, \ref{fig:G345_2}, and \ref{fig:comparison} show how the QA2 images are typically limited to only a portion of the data while ARI-L images provide a more complete overview of the data content. In figure \ref{fig:G345_1}, we show that the QA2 package includes images for 2 of the 13 spectral windows observed for an O-type young stellar object (G345.49+1.47) to study kinematic structures in dense, warm gas associated with star formation (Cesaroni et al. 2017, Moscadelli et al. 2019, Maud et al 2019). The complete spectral coverage of the ARI-L cubes (see Fig. \ref{fig:G345_2}) shows the rich chemistry of this object. The kinematic information from the line emission can be used to investigate the properties of Keplerian disks or outflows and to constrain the mechanisms of massive star formation. Figure \ref{fig:comparison} presents the spectral scan of a region of the Central Molecular Zone of the Milky Way. The QA2 process verified the achievement of PI requirements only for a portion of each spectral windows, while the ARI-L spectra cover the entire frequency range. These cases are examples of the improved information that the ARI-L images offer for spectral analysis.

ARI-L also offers unprecedented easy access to data that can be used for the investigation of light curves, spectral behaviour, variability and morphological analysis of the ALMA calibrators. They are typically blazars (Bonato et al. 2018, 2019), a class of AGNs characterized by strong variability over a broad range of time scales in all wavebands (e.g. Ulrich et al. 1997; Webb 2006; Aharonian et al. 2007, Tavecchio et al. 2011, Abdo et al. 2010; Galluzzi et al. 2017). The entire non-thermal continuum is believed to originate mainly from a relativistic jet pointing close to our line of sight. ARI-L images of calibrators provide an extraordinary amount of multi-epoch and multi-frequency data in the poorly explored mm/sub-mm spectral region that can shed light on the physical processes in action, such as particle acceleration and emission mechanisms, relativistic beaming, the origins of flares, and the size, structure, and location of the emitting regions.

\begin{figure}
  \centering
  \includegraphics[trim=0 5cm 0 5.5cm, clip, width=8cm]{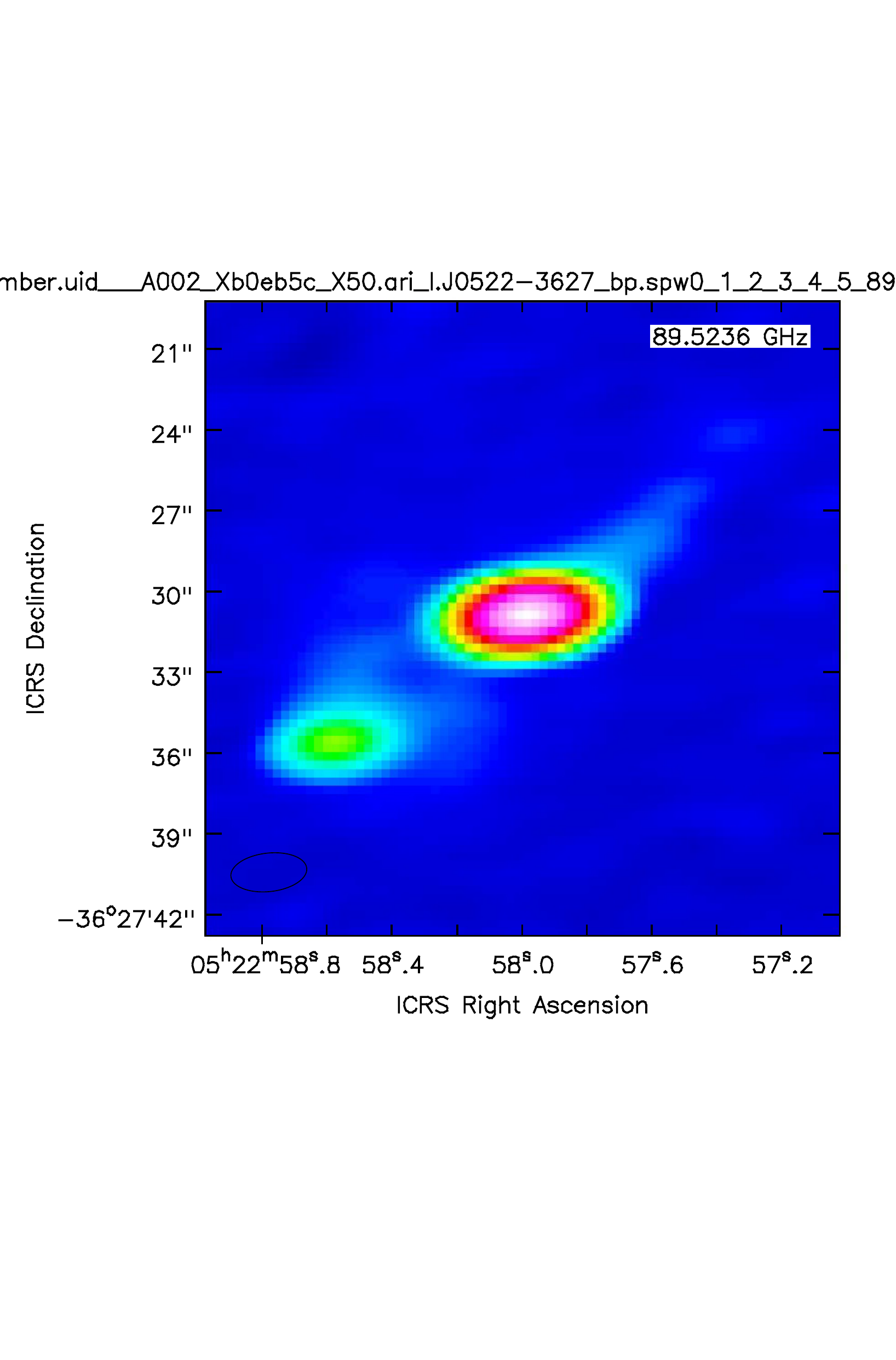}
  \includegraphics[trim=0 5cm 0 5cm, clip, width=8cm]{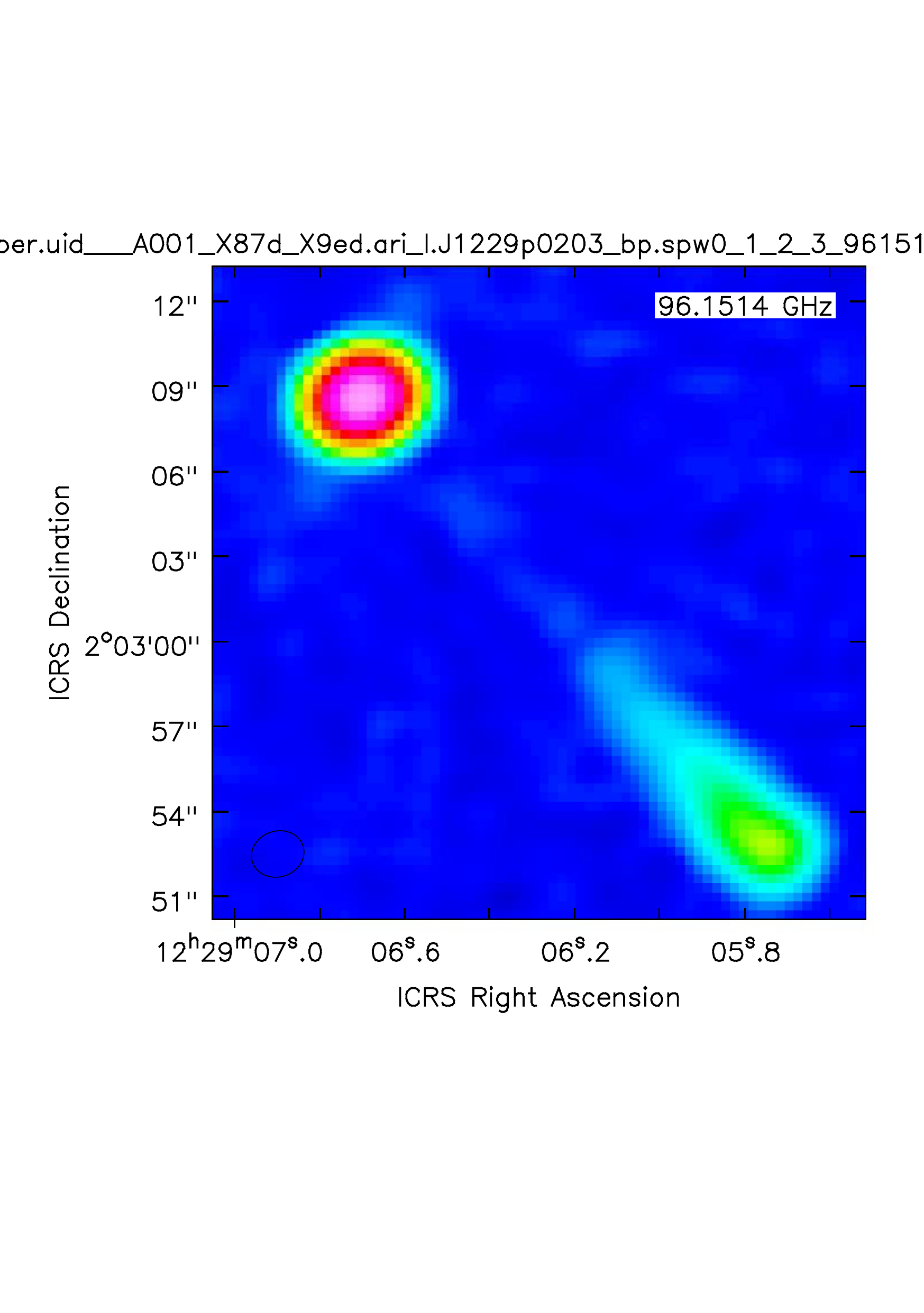}
\caption{ARI-L images of PKS 0521-365 (top) observed in project 2015.1.00503.S MOUS uid://A002/Xb0eb5c/X50 and 3C273 (bottom) observed in project 2016.1.01340.S MOUS uid://A001/X87d/X9ed.}
  \label{fig:calibrators}
\end{figure}

\begin{figure*}
  \centering
\begin{tikzpicture}

\node (img) at (2cm,9cm) {\includegraphics[trim=0cm 5cm 0cm 5cm,clip,width=6cm]{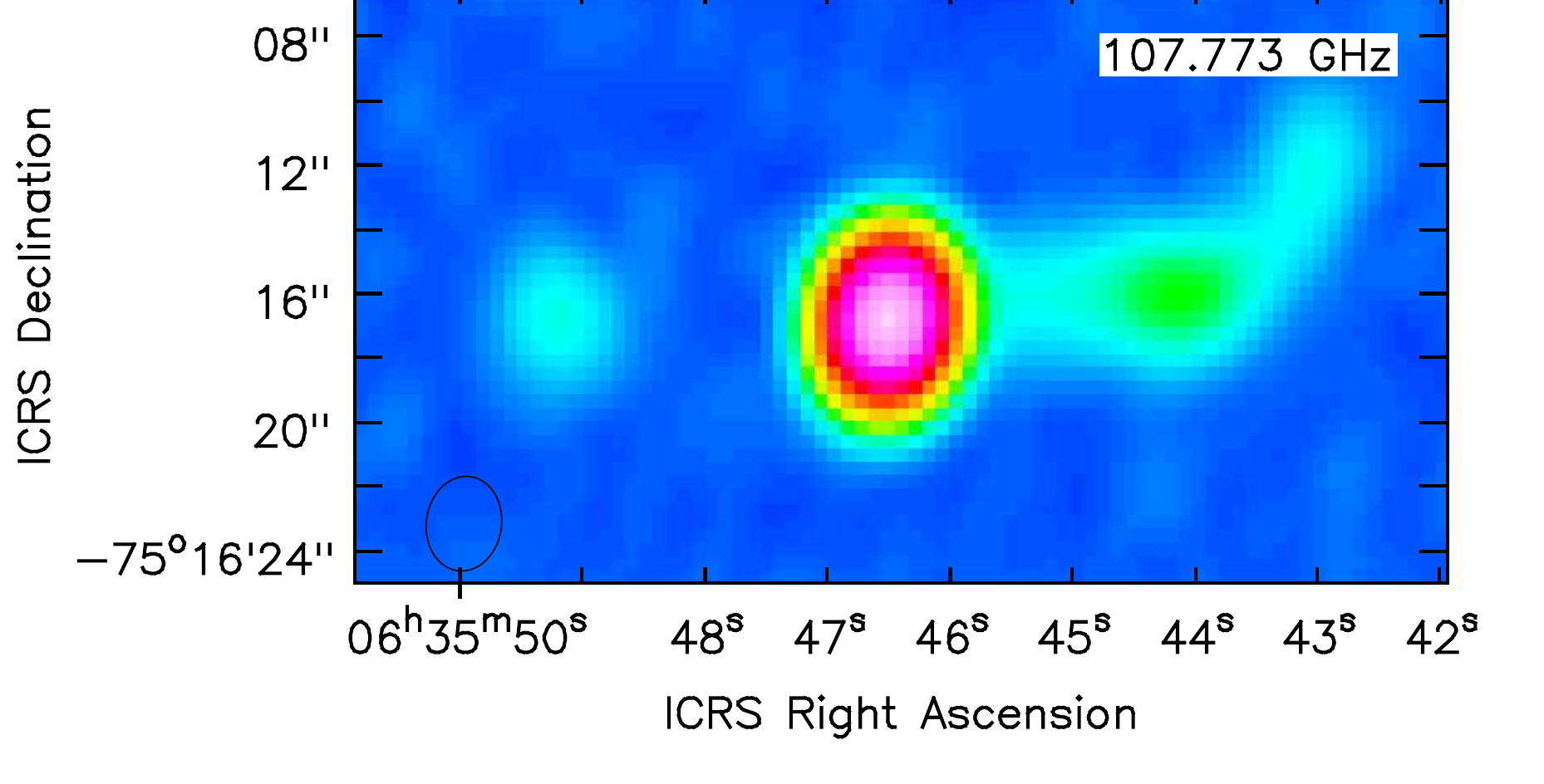}};
\node (img) at (-3cm,9cm) {\includegraphics[trim=0cm 5cm 0cm 5cm,clip,width=6cm]{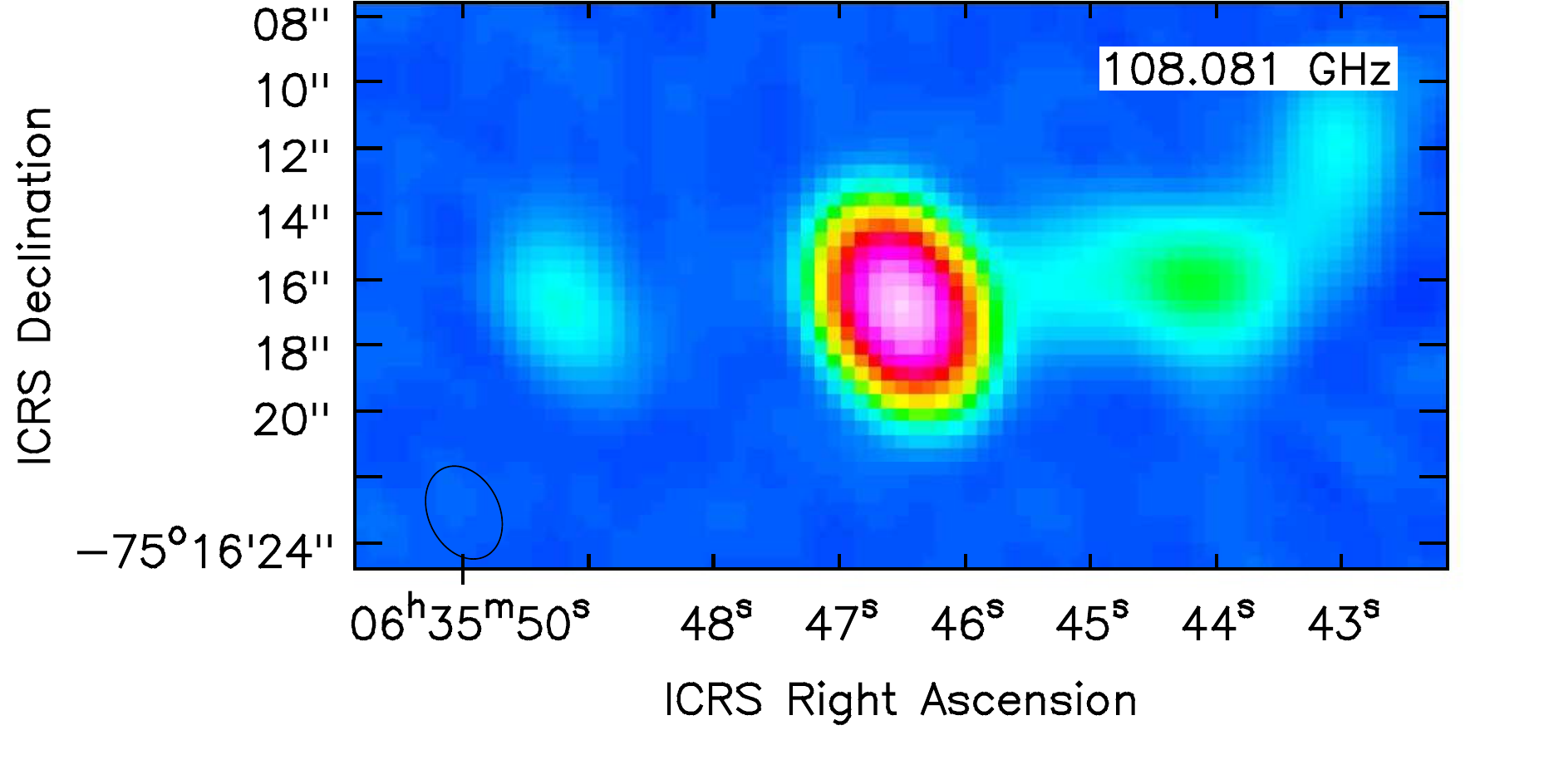}};
\node (img) at (-8cm,9cm) {\includegraphics[trim=0cm 5cm 0cm 5cm,clip,width=5.9cm]{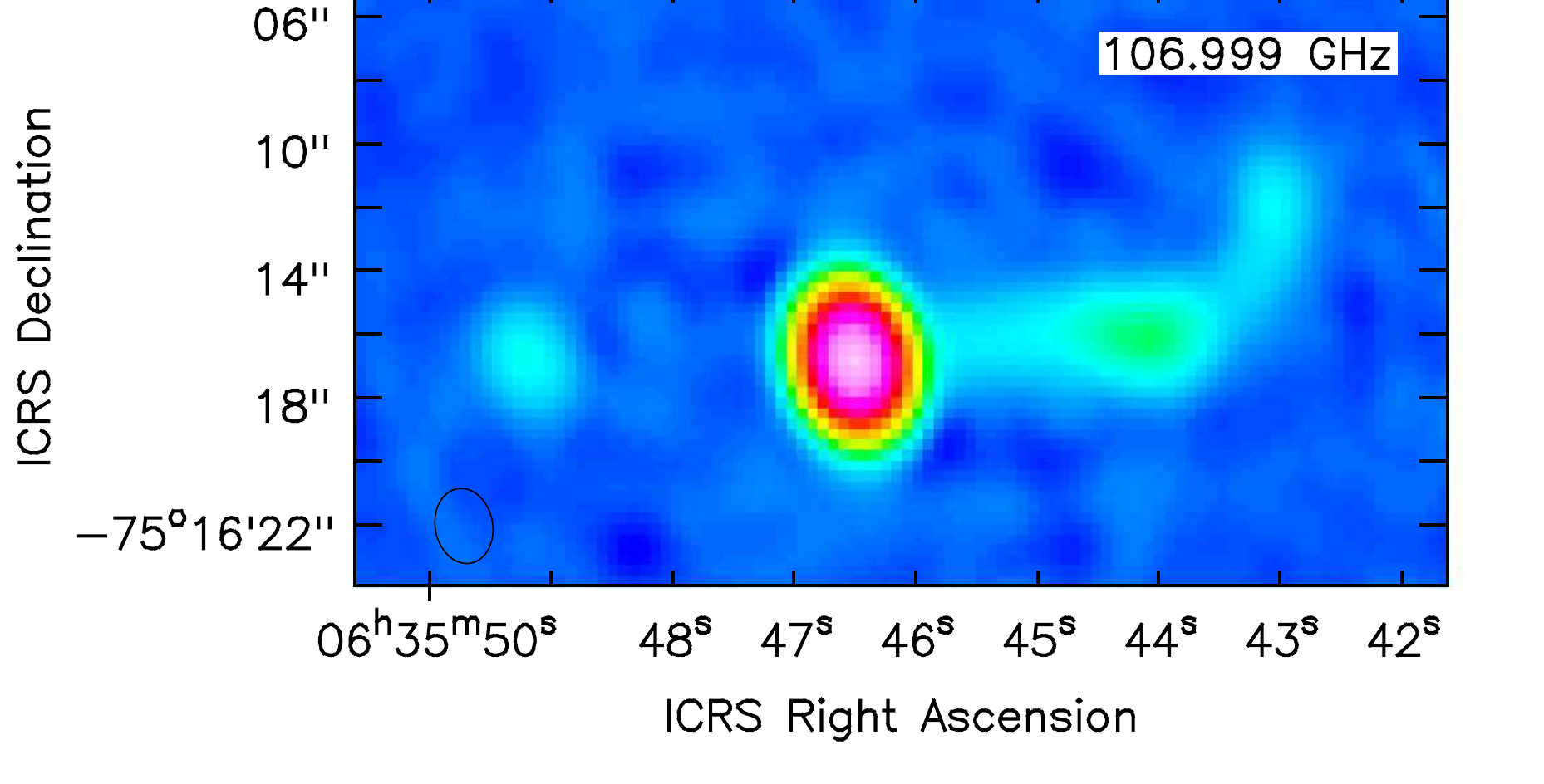}};

\node [draw, fill=white] at (-8.5cm,10cm){2016-01-07};
\node [draw, fill=white] at (-3.5cm,10cm){2016-01-12};
\node [draw, fill=white] at (1.5cm,10cm){2016-01-24};

\node (img) at (2cm,6cm) {\includegraphics[trim=0cm 5cm 0cm 5cm,clip,width=6cm]{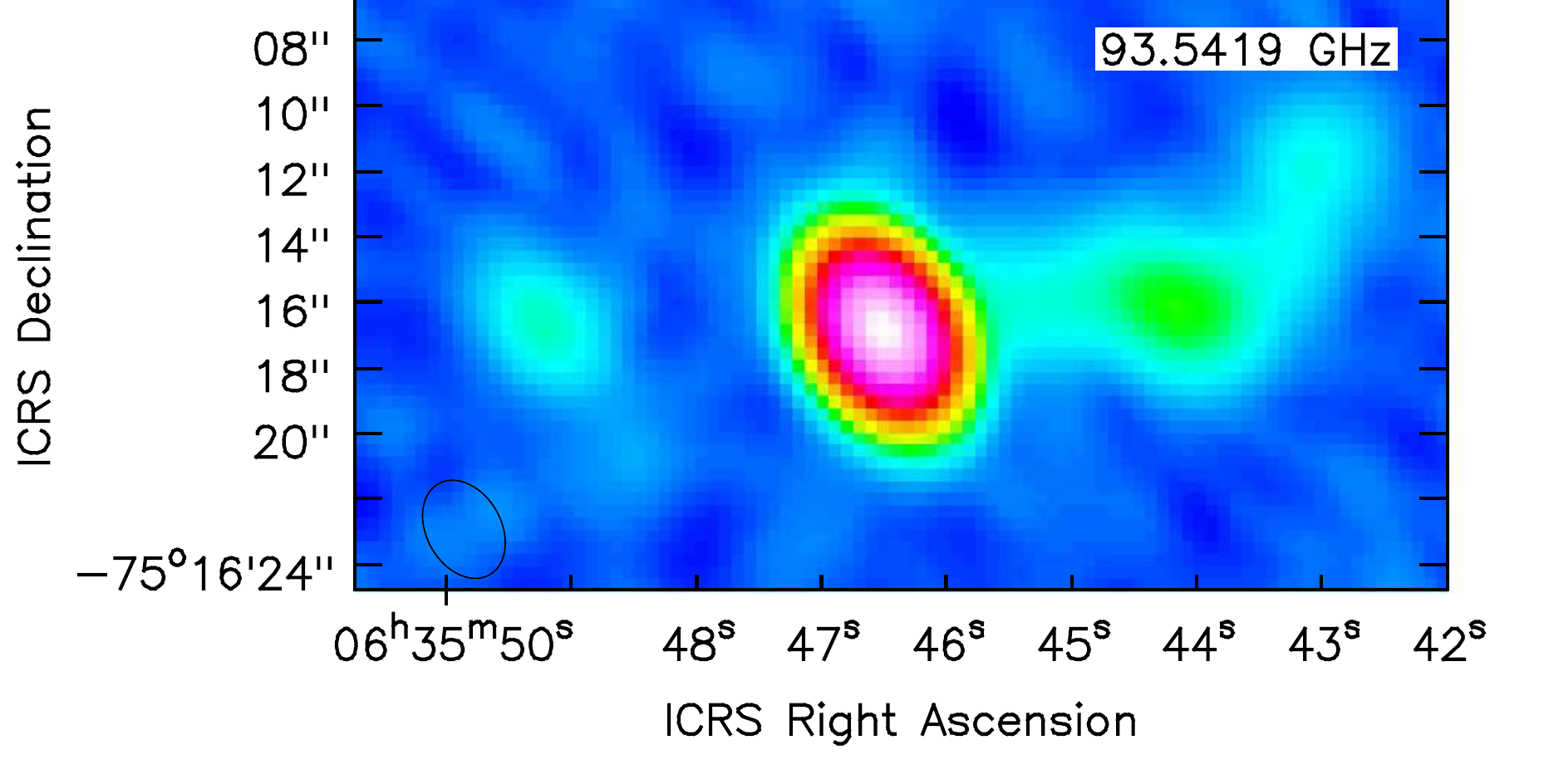}};
\node (img) at (-3cm,6cm) {\includegraphics[trim=0cm 5cm 0cm 5cm,clip,width=6cm]{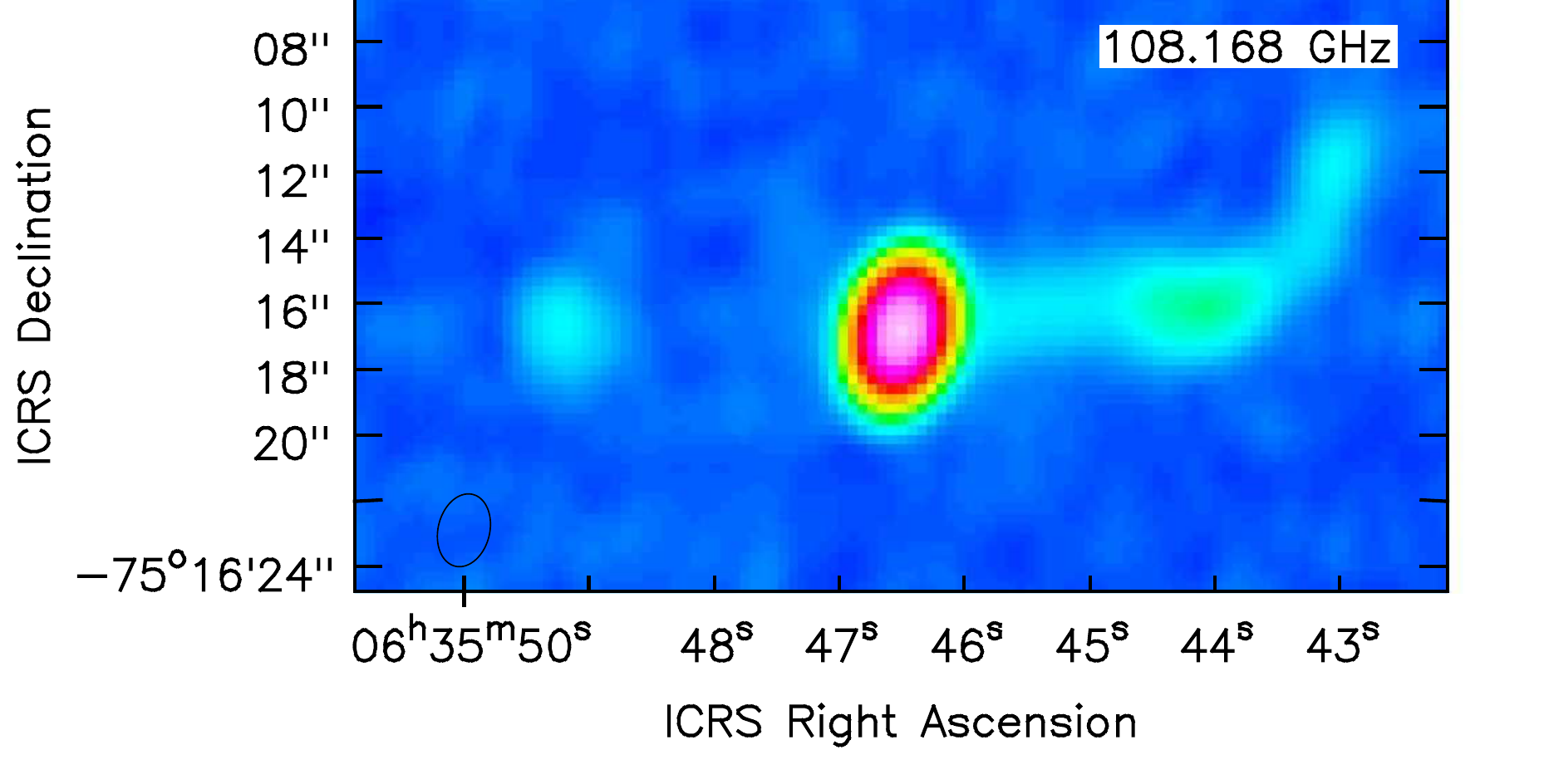}};
\node (img) at (-8cm,6cm) {\includegraphics[trim=0cm 5cm 0cm 5cm,clip,width=6cm]{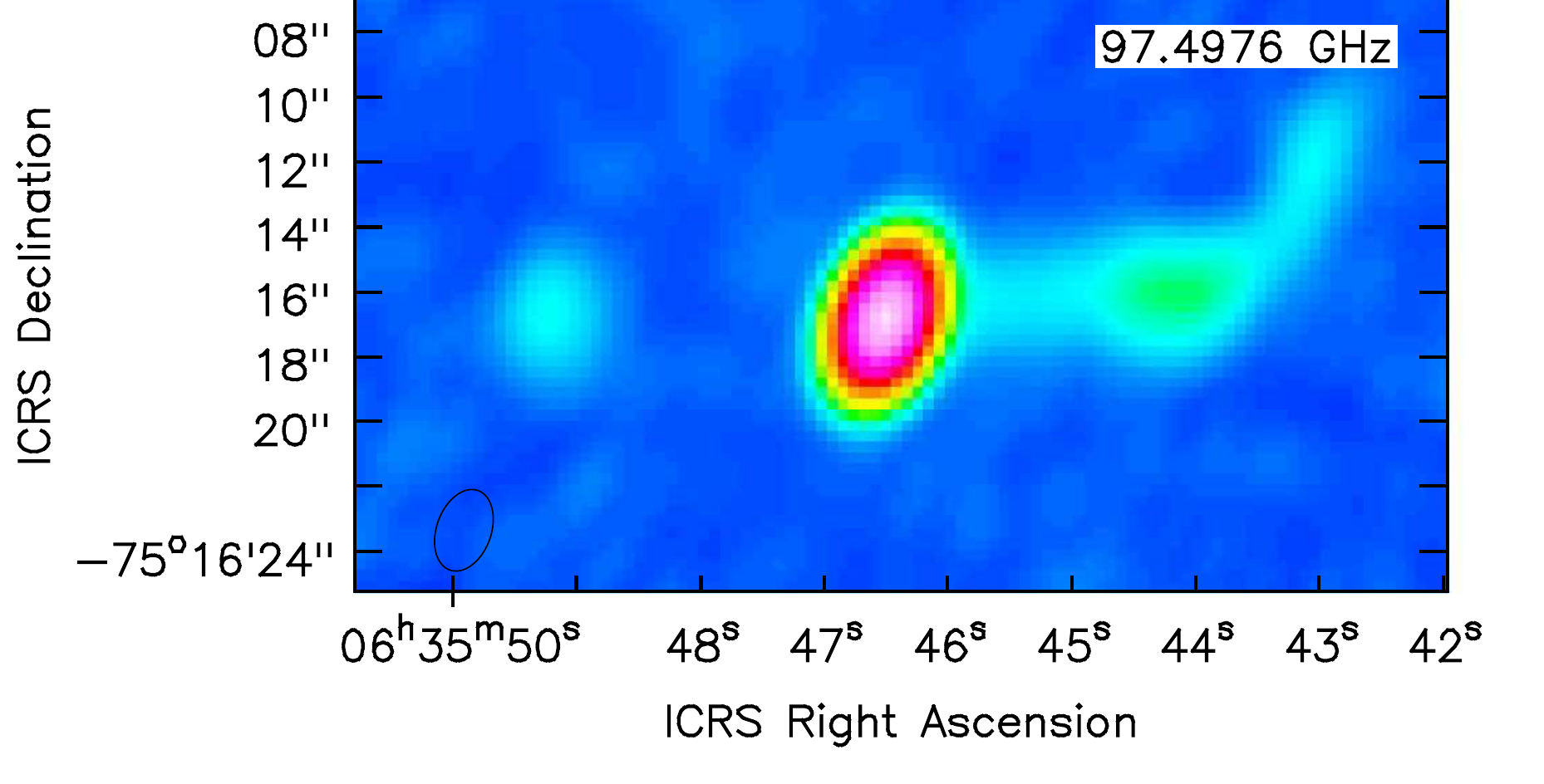}};

\node [draw, fill=white] at (-8.5cm,7cm){2016-03-10};
\node [draw, fill=white] at (-3.5cm,7cm){2016-03-26};
\node [draw, fill=white] at (1.5cm,7cm){2016-04-04};

\node (img) at (2cm,3cm) {\includegraphics[trim=0cm 5cm 0cm 5cm,clip,width=6cm]{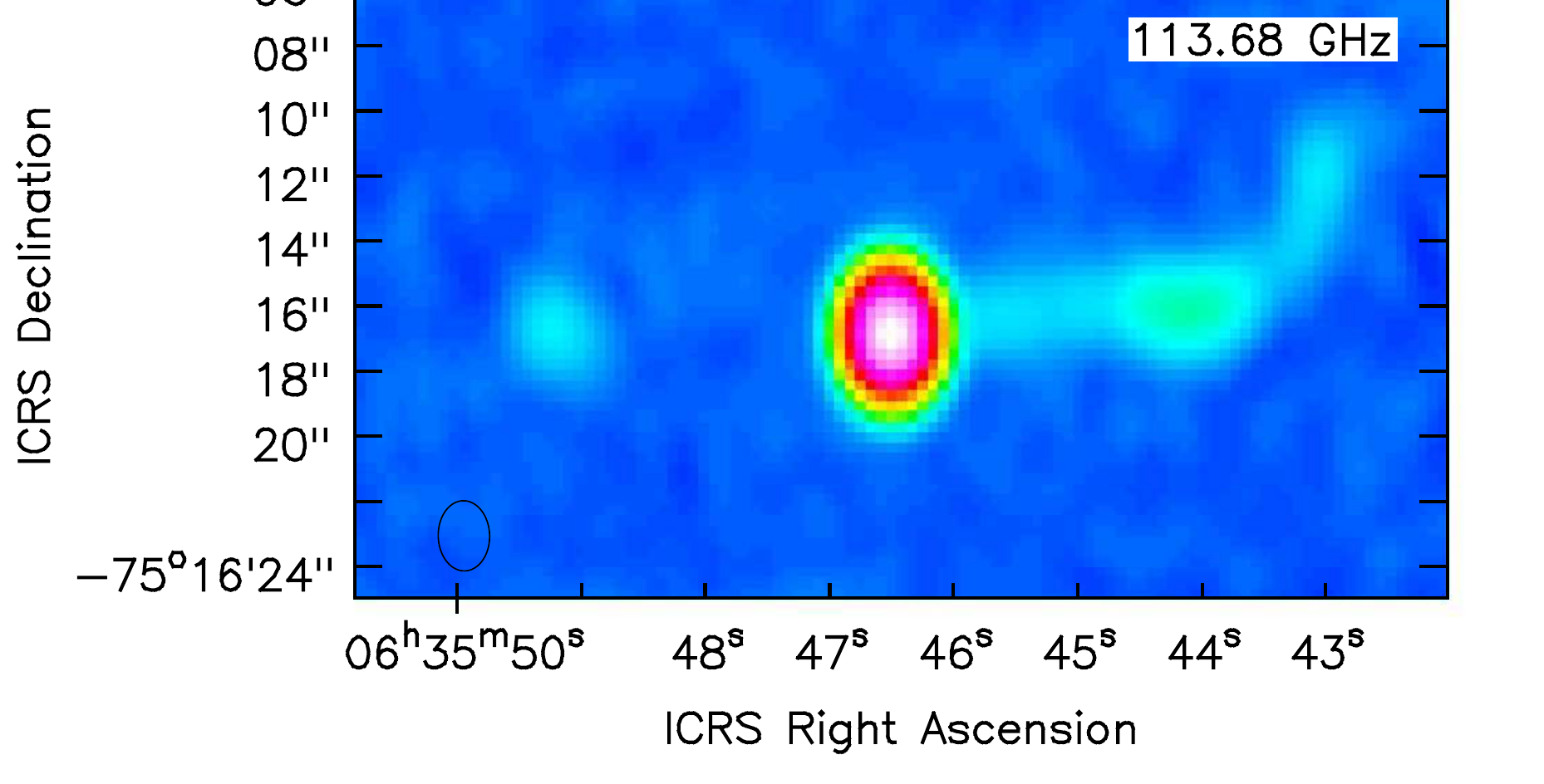}};
\node (img) at (-3cm,3cm) {\includegraphics[trim=0cm 5cm 0cm 5cm,clip,width=6cm]{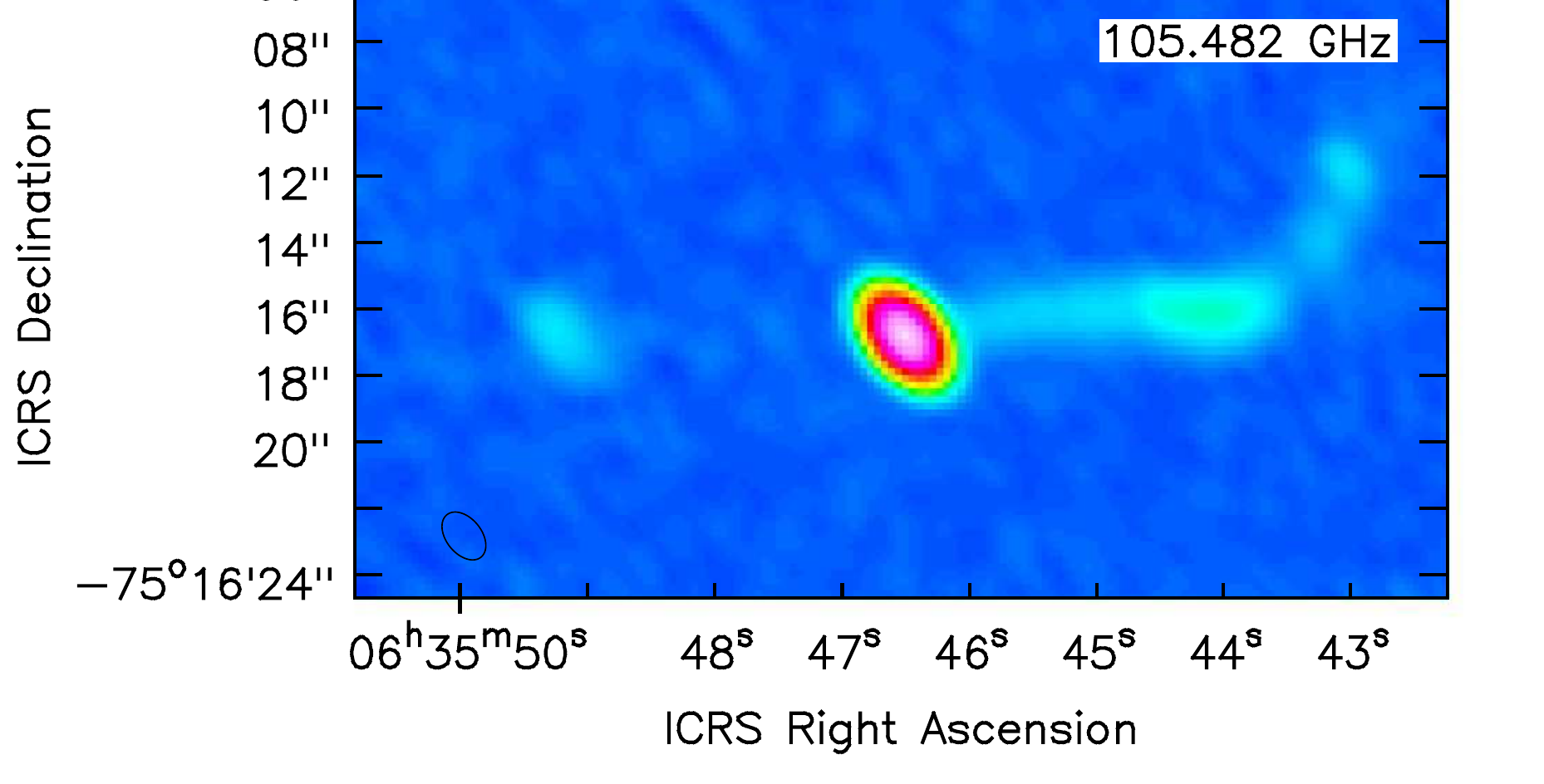}};
\node (img) at (-8cm,3cm) {\includegraphics[trim=0cm 5cm 0cm 5cm,clip,width=6cm]{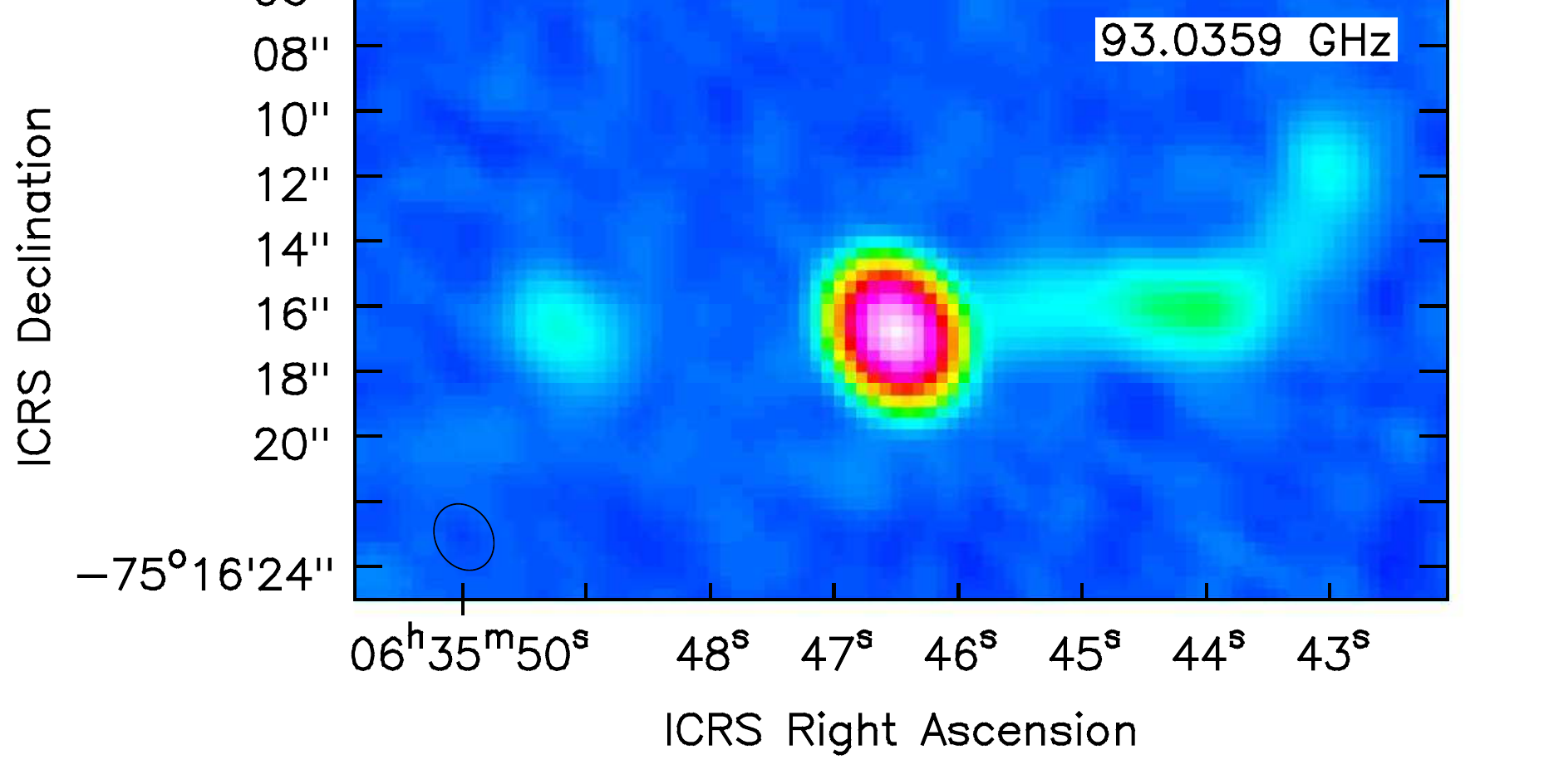}};

\node [draw, fill=white] at (-8.5cm,4cm){2016-04-16};
\node [draw, fill=white] at (-3.5cm,4cm){2016-06-18};
\node [draw, fill=white] at (1.5cm,4cm){2016-12-31};

\node (img) at (2cm,-0.5cm) {\includegraphics[trim=0cm 0cm 1cm 0cm,clip,width=5cm, height=4cm]{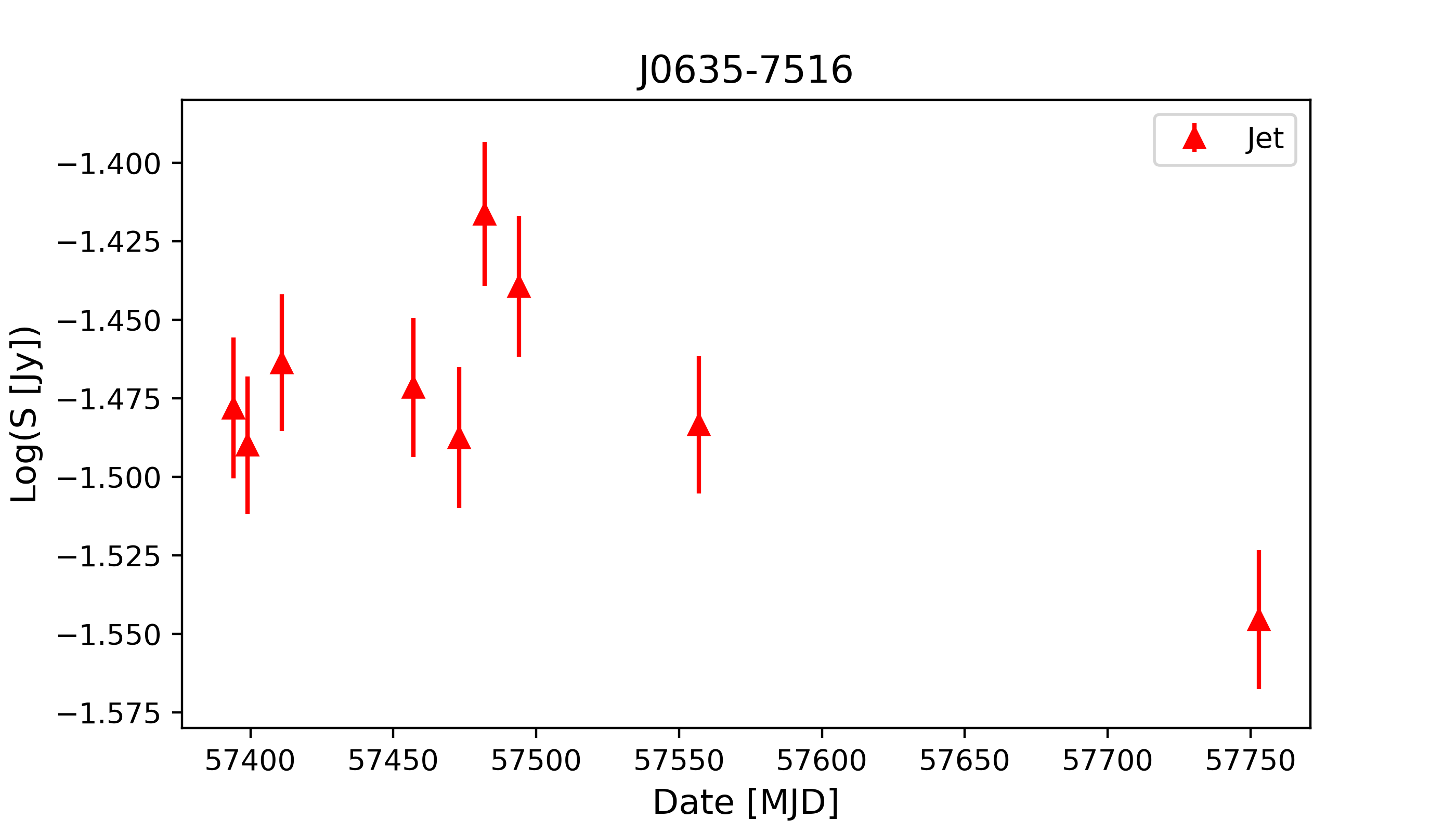}};
\node [draw, fill=white] at (2cm,1cm){$\ \ Jet\ \ $};
\node (img) at (-3cm,-0.5cm) {\includegraphics[trim=0cm 0cm 1cm 0cm,clip,width=5cm, height=4cm]{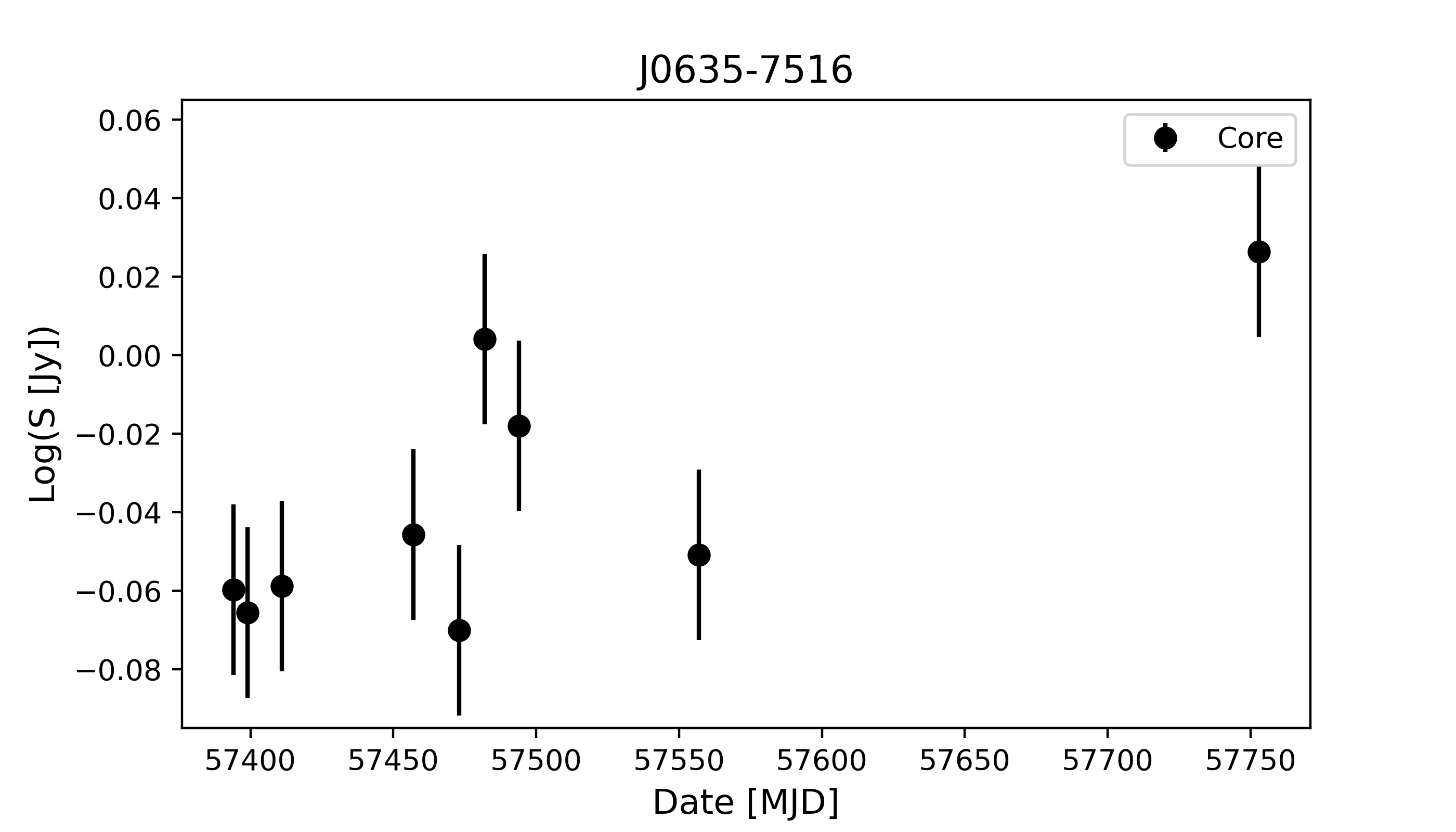}};
\node [draw, fill=white] at (-3cm,1cm){$\ \ Core\ \ $};
\node (img) at (-8cm,-0.5cm) {\includegraphics[trim=0cm 0cm 1cm 0cm,clip,width=5cm, height=4cm]{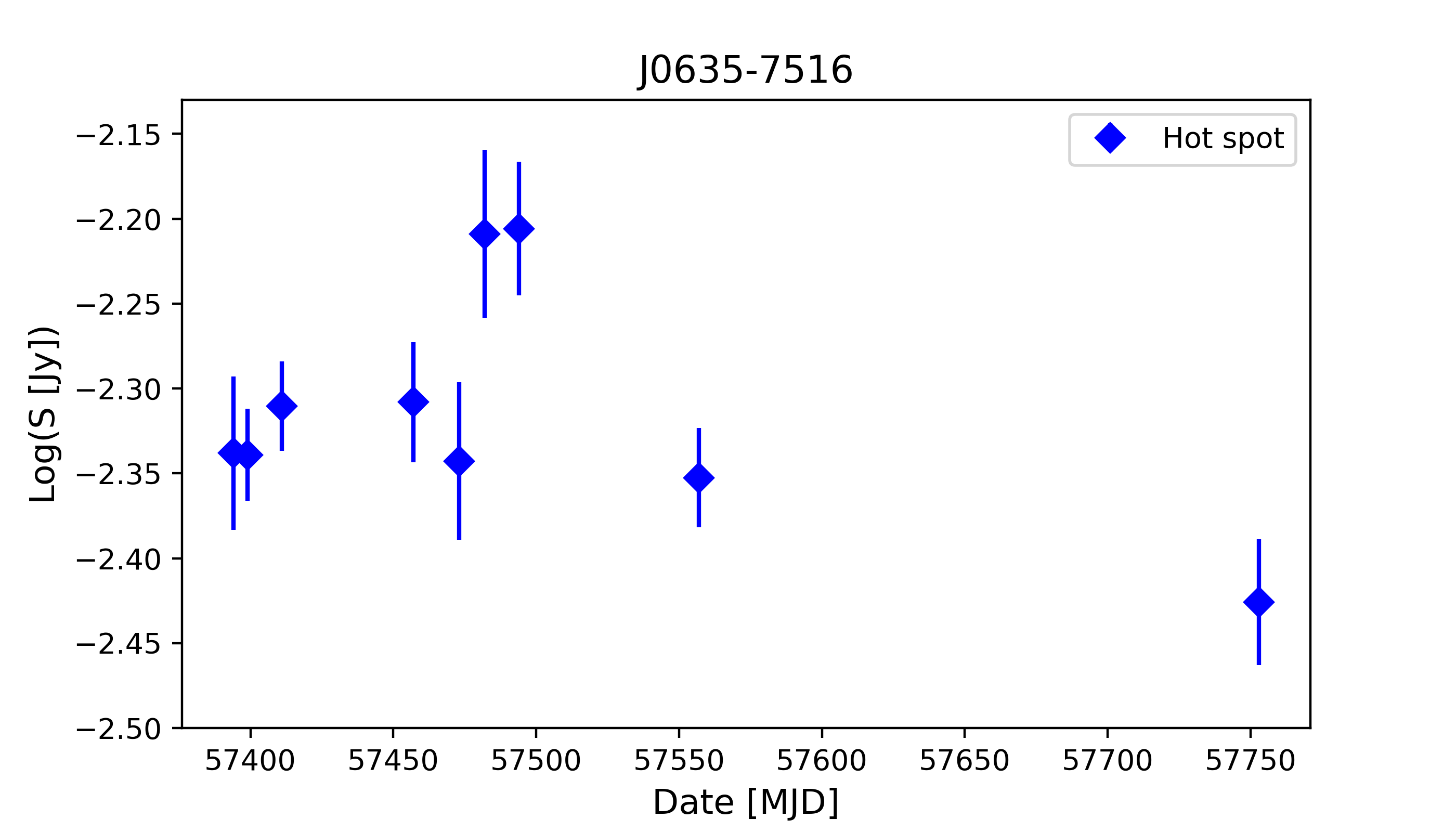}};
\node [draw, fill=white] at (-8cm,1cm){$Hot spot$};

\end{tikzpicture}
\caption{ARI-L images of 105 GHz observations of PKS 0637-752 from Cycle 3 projects observed in 2016 (with labels indicating the observation date) and light curves of the hot-spot, core, and jet regions (with error bars that include a calibration error equal to 5\% of the flux density). The various angular resolutions of the data available in the archive make it possible to investigate the source structure. This blazar is strongly core-dominated, and the second brightest component is only 2 orders of magnitude fainter than the core in this band.  The projects used for this figure are 2015.1.01046.S, 2015.1.01195.S, 2015.1.00190.S 2015.1.00204.S, 2015.1.00196.S, 2015.1.01388.S, 2015.1.00697.S, and 2016.1.00193.S.}
  \label{fig:0635_all}
\end{figure*}
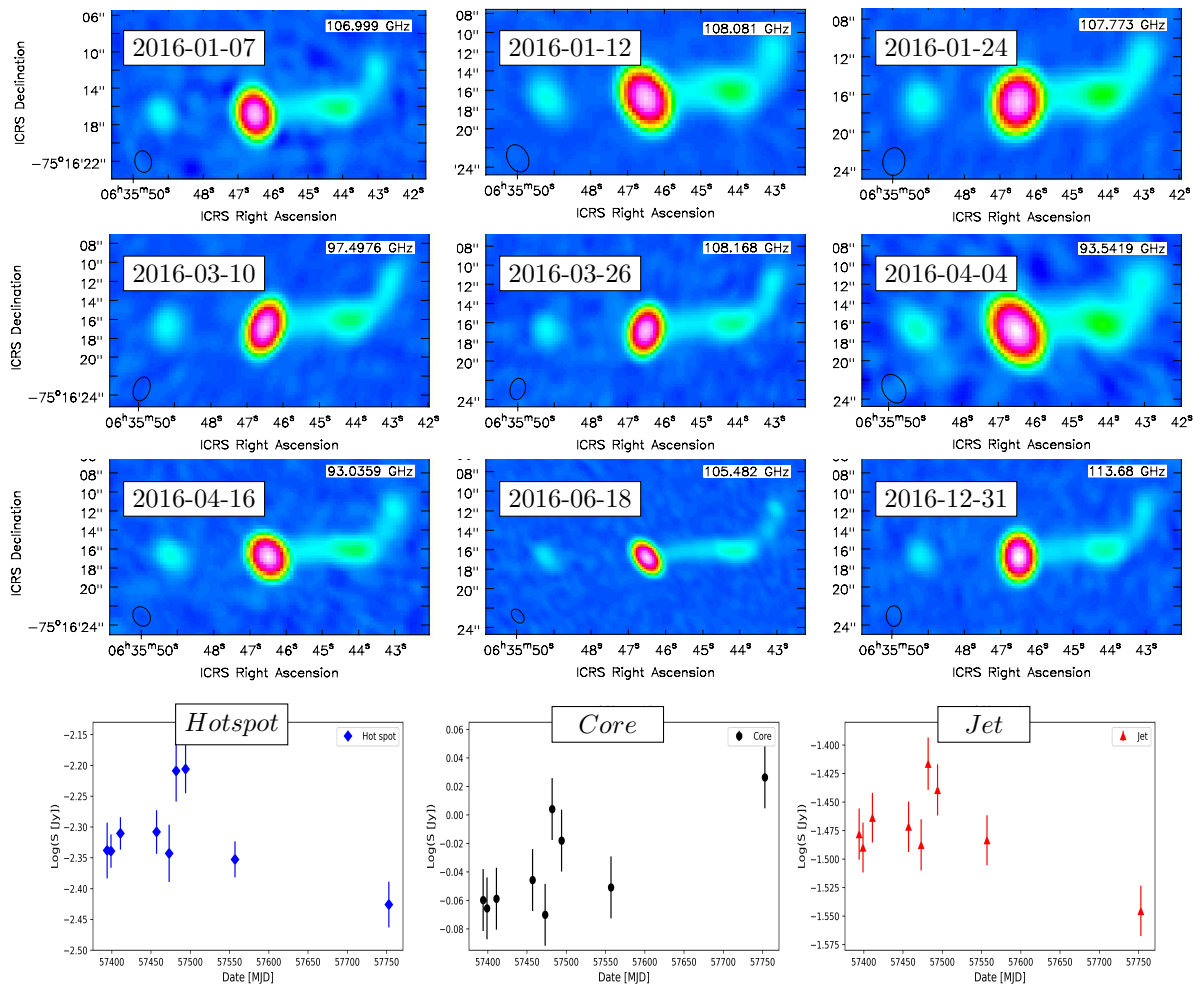

\begin{figure*}
  \centering
  \includegraphics[width=\textwidth]{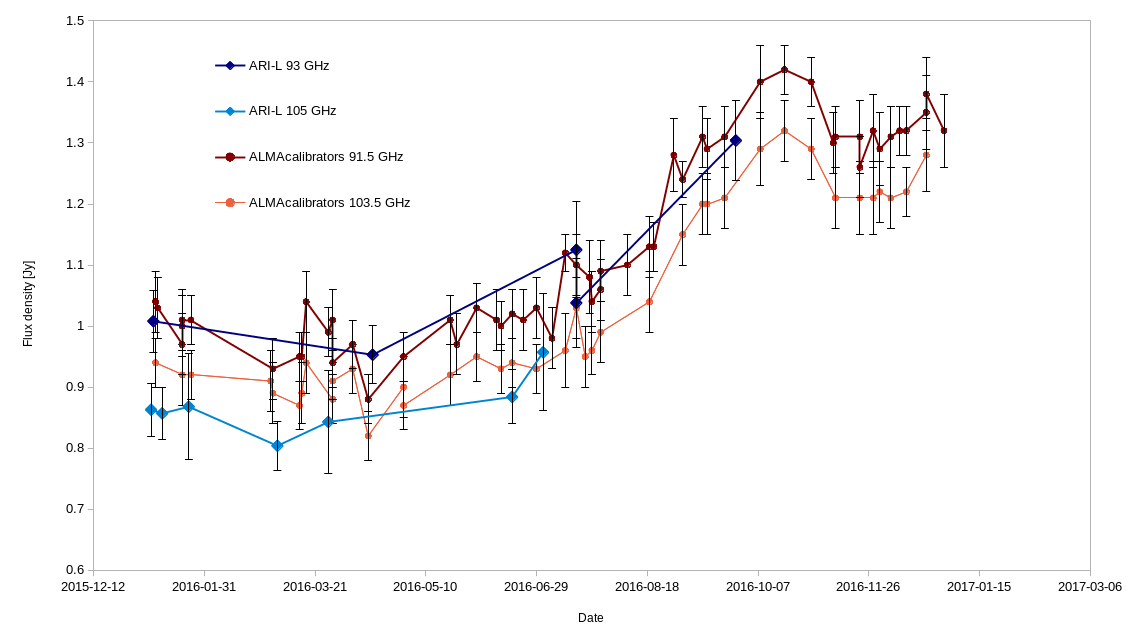}
\caption{The 2016 light curves extracted from the core region of PKS 0637-752. Data from the ARI-L images at $\sim$93 and $\sim$105 GHz are shown as blue and cyan diamonds, while the ALMA Calibrator catalogues values available at 91.5 and 103.5 GHz are shown as red and orange squares. }
  \label{fig:0635_lc}
\end{figure*}

Figures~\ref{fig:calibrators}, \ref{fig:0635_all}, and \ref{fig:0635_lc} show some of the capabilities offered by ARI-L images for studying the morphology and variability of blazars. We present images for three ALMA calibrators with well-known resolved structure, namely PKS0521-365, 3C273, and PKS0637-752 (see also for comparison Meyer et al. 2017 and Agliozzo et al. 2017 for PKS 0637-752, Hovatta et al. 2018 for 3C273, and Liuzzo et al. 2015 for PKS 0521-365).

For PK0637-752 we present some of the ARI-L continuum images for Cycle 3 projects observed in 2016 with average frequency of $\sim$105 GHz that have been ingested into the archive thus far. We extracted light curves from the core, jet and hot-spot regions that are similar to what could be used for investigating the evolution of variability along the source substructures. We also compared the core region light curves at $\sim$105 GHz and at $\sim$93 GHz with the data available in similar frequency ranges in the ALMA Calibrator Manual\footnote{https://almascience.eso.org/alma-data/calibrator-catalogue}, which provides data with very good time coverage but does not provide images.

\begin{figure*}
  \centering
   \includegraphics[width=0.3\textwidth]{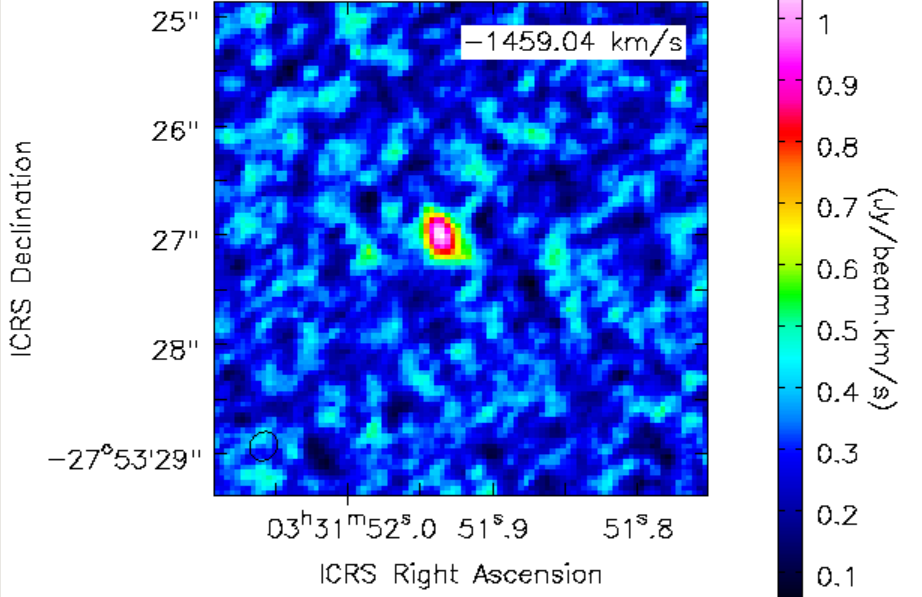}
  \includegraphics[width=0.3\textwidth]{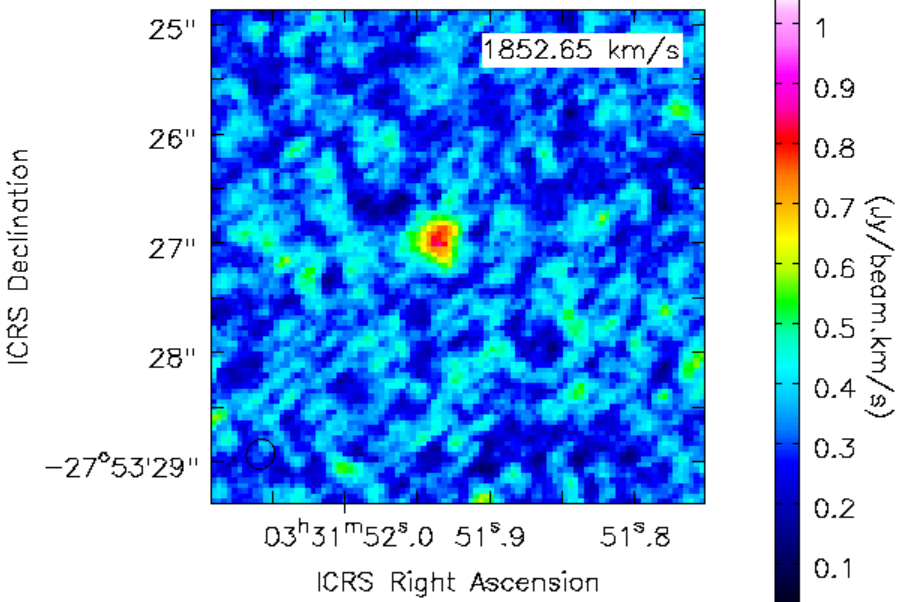}
  \includegraphics[width=0.3\textwidth]{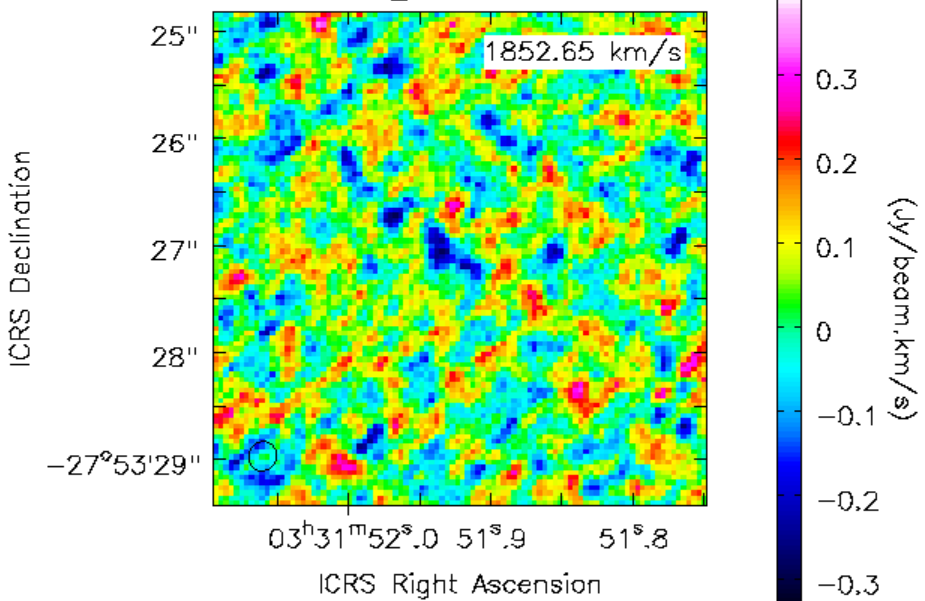}
  \includegraphics[width=\textwidth, height=4cm, clip]{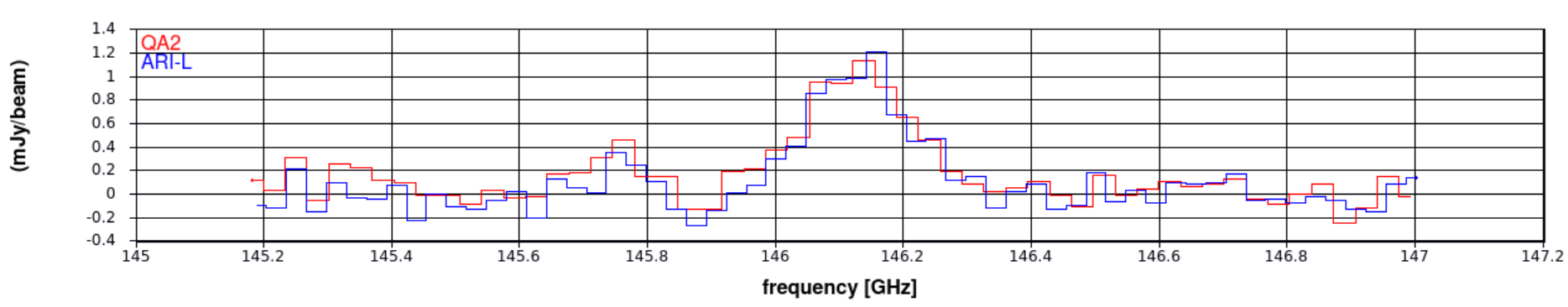}
  \caption{Zeroth moments of the brightness distribution (corresponding to integrated brightness) of the CO(5-4) line for the dusty obscured galaxy XID42 (D'Amato et al. 2020) obtained exploiting the QA2 (top left) and the rebinned ARI-L  (center) products for MOUS uid://A001/X340/X2c of the project 2015.1.01205.S. For comparison we present also the the difference between the zeroth moments (top right), and the comparison of the line spectra}
  \label{fig:XID42}
\end{figure*}

The ARI-L product images are produced at the native spectral resolution. Quite often, spectral features cannot immediately be detected within ALMA data in their native spectral resolution, in which case the CASA task IMREBIN could be used to enhance spectral line signals above the noise. IMREBIN, in fact, averages along the spectral axis of a cube by a given factor without considering any difference among the signals in the different channels \footnote{https://casa.nrao.edu/docs/taskref/imrebin-task.html}.

In an interferometric dataset, visibilities for the same baseline B at different frequencies, $\nu$, observe emission on different angular scales, $\theta$, according to the relation $\theta\propto B c/\nu $, where $c$ is the speed of light.

On the one hand, when generating images at their native spectral resolution, the different channels sample different angular scales, and the difference in the angular scales is related to the difference in the frequencies sampled by the channels. When contiguous channels are averaged together in the rebinning task, channels with information about emission on different scales are averaged together. Furthermore, different channels correspond to different synthesized beam sizes, so even a point source appears larger at longer wavelength. (Also, IMREBIN does not apply any smoothing before averaging.)

On the other hand, if visibilities from different channels are used in the generation of an image with lower spectral resolution, then they contribute to the definition of larger channels within the image cube that combine the signals on all the sampled angular scales according to the chosen weighting scheme.

The difference between the high-resolution rebinned products and the low-resolution imaged products is more significant in case the Fourier transform of the target brightness distribution is rapidly variable or if the overall coverage of the Fourier domain is sparse, as the products might be more sensitive to different visibility sampling.

It is always recommended to reproduce the images when specific spectral resolutions are needed. Nonetheless, we investigated the differences between a rebinned ARI-L image to an image created with {\sc clean} or {\sc tclean} using a lower spectral resolution.  For this analysis, we used observations of the CO(5-4) line in the spectrum of an obscured AGN at redshift 2.94 (D'amato et al. 2020, see fig. \ref{fig:XID42}).  CO lines are well-recognized tracers of molecular gas and indirectly of star formation at any redshift (Kennicutt 1998, Carilli et al. 2013), with the highest transitions more significantly excited in denser and more energetic environments like those surrounding the supermassive black holes in AGN. There is increasing evidence of common evolutionary behaviour between star formation and BHs, with both cosmic densities peaking at redshift 2-3 (Madau \& Dickinson 2014, Gruppioni et al. 2013).

According to models of BH-star formation in-situ co-evolution (Lapi et al. 2018, Pantoni et al. 2019), dusty star forming galaxies should span regions with sizes of 1-5 kpc at the peak of cosmic evolution.  This corresponds to 0.12-0.6 arcseconds, which is easily achieved by ALMA in its intermediate configurations, even at the lowest frequencies.  In these data, the galaxies might appear barely resolved (Pantoni et al. in prep, D'Amato et al. 2020).

CO lines in galaxies at the peak of cosmic star formation might be several hundreds of km/s in width, corresponding to many tenths of MHz at $\sim$100GHz. This would span a few tens of channels in ALMA data, even in data with low spectral resolutions. However, the presence (or the absence) of a spectral line in a target might determine its inclusion in selected samples, so it is important to quickly establish if a line could be detected from archived images.

Fig.\ref{fig:XID42} presents the zeroth moment of the CO(5-4) spectral line for XID 42 as obtained by imaging the target at the spectral resolution of 70 km/s, which is what was available from the archived QA2 products (i.e. imaged by QA2 analysts). For comparison, the ARI-L image of the same target (calibrated with the same script, but imaged with the Imaging Pipeline) was rebinned by a factor 4 along the frequency axis to match its spectral resolution to the QA2 data as closely as possible. The rms noise in the zeroth moment images is 0.31 and 0.34 (mJy/beam) km/s respectively for the QA2 and rebinned ARI-L images. At the position of the source, the maximum absolute difference between the two is 0.28 (mJy/beam) km/s.  When the resulting spectra are also compared, the spectral lines are clearly similar within the noise variation, and the pattern of noise (that, in an interferometric image, is not genuinely Poissonian but determined by the visibility domain coverage that is common between the two images) is similar.

Hence, the detection of a CO line in the rebinned ARI-L image is indicative of the presence of the line in images produced with {\sc clean} or {\sc tclean} using coarser resolution settings.

\begin{figure}
  \centering  \includegraphics[width=0.45\textwidth]{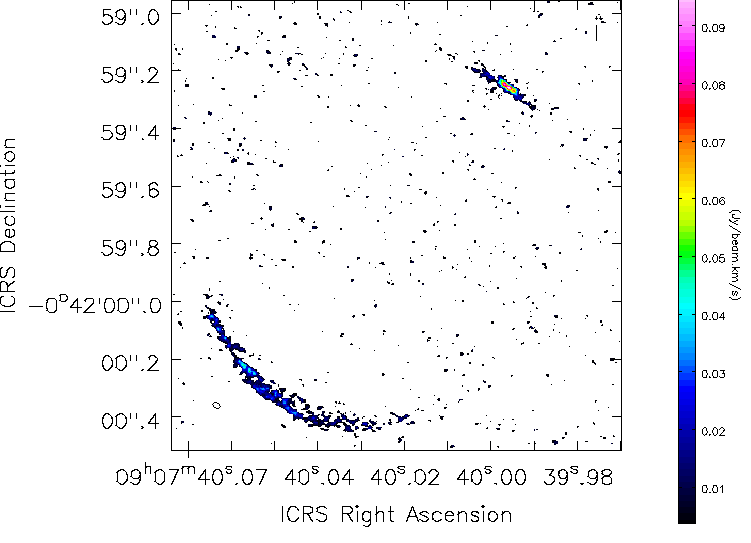}

  \includegraphics[width=0.45\textwidth]{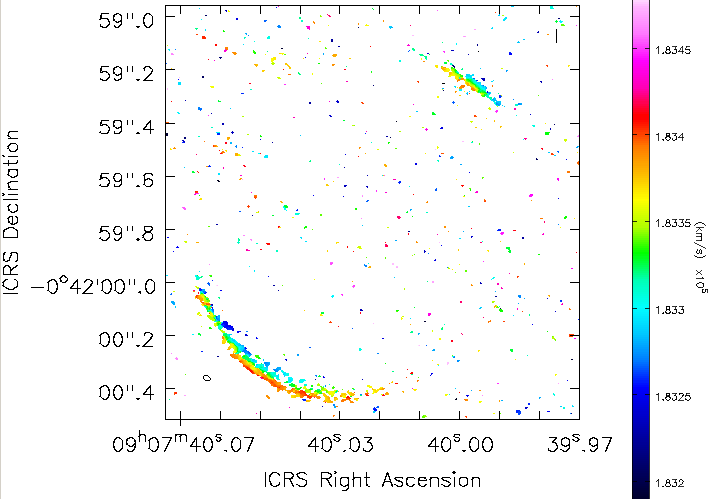}

  \includegraphics[width=0.45\textwidth]{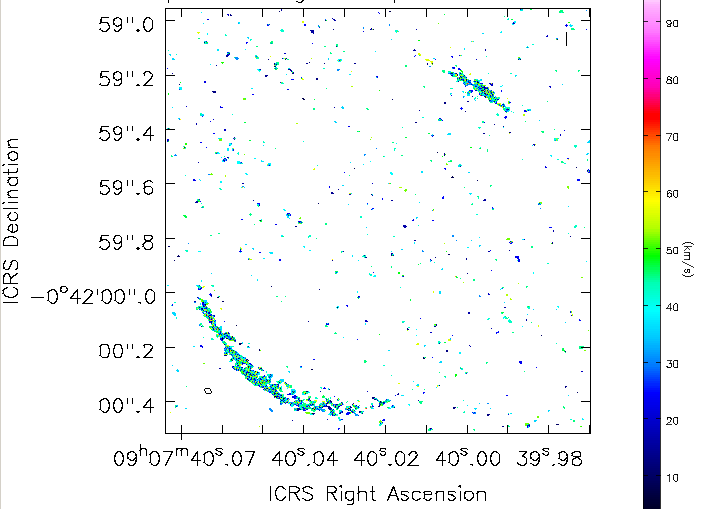}
  \caption{Zeroth, first, and second moment images (corresponding to integrated brightness, velocity distribution and velocity dispersion; top to bottom) of the CO(6-5) line for the lensed galaxy SDP9 obtained using the ARI-L products for MOUS uid://A001/X2de/X28 of the project 2015.1.00415.S (for more info see Massardi et al. 2018).}
  \label{fig:sdp9}
\end{figure}

Similarly, first moment maps of spectral line distributions produced from ARI-L images could be informative of the kinematic behaviour of the objects. Figure \ref{fig:sdp9} shows the moment images corresponding to the integrated brightness, velocity distribution, and velocity dispersion that have been obtained from the archived ARI-L image of the CO(6-5) line of the lensed galaxy SDP9.  These images clearly show the presence of a disk or an outflow in this $z=1.57$ lensed galaxy at the same significance levels that members of our collaboration achieved a few years ago in a complete analysis of the archived dataset (see fig. 5 in Massardi et al. 2018 for comparison).

\section{Summary}\label{sec:summary}

The Additional Representative Images for Legacy (ARI-L) project is a European Development project for ALMA Upgrade approved by JAO and ESO that officially started in June 2019 and is currently on-going and well-ahead of schedule. The project aims to increase the legacy value of the ASA by bringing the reduction level of ALMA data from Cycles 2-4 close to the level of the more recent Cycles processed with the ALMA Imaging Pipeline. Over a period of three years, the ARI-L project will produce and ingest into the ASA a uniform set of full data cubes and continuum images at native spectral resolution and average weighting scheme, that cover at least 70\% of the observational data from Cycles 2-4 that can be processed with the ALMA Imaging Pipeline.

By the end of the project, we expect to have produced and delivered to the ALMA Science Archive images for at least 2733 MOUS, which corresponds to a minimum of 70\% of the 3476 MOUS that we expect the current version of the Imaging Pipeline could process. As of June 2021, more than 1800 of them have been reprocessed, and more than 150000 images have been returned to the ASA and are available to astronomers. These products can already be retrieved from the ALMA Science Archive as "Externally delivered products" on the Request Handler download page; the ARI-L products are listed directly below the ALMA data.
As of June 2021, there have been more than 146800 downloads of a file tagged "ari-l" from the ASA, either as an individual file or included in packages (ASA staff private communication; a single file can be downloaded more than once).
Furthermore, all the corresponding calibrated measurement sets are stored on the INAF-IA2 facilities and can be requested through the ARI-L project website https://almascience.eso.org/alma-data/aril
where also further details on the project status are available.

The ARI-L cubes and continuum images complement the much smaller QA2-generated archival image products, which cover only a small fraction of the available data for Cycles 2-4. ARI-L imaging products are highly relevant for many science cases and significantly enhance the possibilities for exploiting archival data. They facilitate archive access and data usage for science purposes even for non-expert data miners, they provide a homogeneous view of archival data for better dataset comparisons and download selections, they make the archive more accessible to visualization and analysis tools, and they enable the generation of preview images and plots similar to those that can be created for subsequent Cycles.

\section*{Acknowledgements}

We thank Leonardo Testi, Tony Mroczkowski, Ciska Kemper, Carlos De Breuck, and the ARI-L review committee chaired by Erich Schmid for the useful comments they provided to our project. We thank Sandra Burkutean, Andrea Lapi, Quirino D'Amato, Lara Pantoni, and Giovanni Sabatini for their contribution to the discussion on the reported examples of science cases and to the project. We thank the anonymous Referee for the constructive comments to our manuscript.

This paper makes use of the following ALMA data: ADS/JAO.ALMA\#2016.1.00288.S,\\ ADS/JAO.ALMA\#2016.1.00875.S,\\
ADS/JAO.ALMA\#2015.1.00503.S\\ ADS/JAO.ALMA\#2016.1.01340.S\\ ADS/JAO.ALMA\#2015.1.01046.S,\\ ADS/JAO.ALMA\#2015.1.01195.S,\\ ADS/JAO.ALMA\#2015.1.00190.S,\\ ADS/JAO.ALMA\#2015.1.00204.S,\\ ADS/JAO.ALMA\#2015.1.00196.S,\\  ADS/JAO.ALMA\#2015.1.01388.S,\\ ADS/JAO.ALMA\#2015.1.00697.S,\\ ADS/JAO.ALMA\#2016.1.00193.S,\\ ADS/JAO.ALMA\#2015.1.01205.S, and\\
ADS/JAO.ALMA\#2015.1.00415.S as examples of the Cycle 2--4 projects included in the ARI-L processing lists.
ALMA is a partnership of ESO (representing its member states), NSF (USA) and NINS (Japan), together with NRC (Canada), MOST and ASIAA (Taiwan), and KASI (Republic of Korea), in cooperation with the Republic of Chile. The Joint ALMA Observatory is operated by ESO, AUI/NRAO and NAOJ.

MM acknowledges the support from grant PRIN MIUR 2017 - 20173ML3WW\_001.

MB acknowledges support from the Ministero degli Affari Esteri della Cooperazione Internazionale - Direzione Generale per la Promozione del Sistema Paese Progetto di Grande Rilevanza ZA18GR02.

\section*{Data Availability}

All the input raw data and the ARI-L product already made public are available through the ALMA Science Archive query interface at https://almascience.eso.org/asax/.



\end{document}